\documentclass[a4paper,11pt]{article}
\pdfoutput=1 

\usepackage{jheppub} 
\makeatletter
\DeclareRobustCommand*{\bfseries}{%
  \not@math@alphabet\bfseries\mathbf
  \fontseries\bfdefault\selectfont
  \boldmath
}
\makeatother

\usepackage{calc}
\usepackage{cleveref}

\usepackage{rotating}
\usepackage[english]{babel}

\usepackage{graphicx}
\usepackage{float}

\usepackage{subcaption}

\usepackage{amsmath}
\usepackage{amssymb}
\usepackage{amsthm}
\usepackage{latexsym}

\usepackage{slashed}

\usepackage{dcolumn}
\RequirePackage[normalem]{ulem}

\restylefloat{figure}

\newcommand{\newc}{\newcommand*}

\long\def\begincomment#1\endcomment{%
        \begingroup\sf\baselineskip12pt#1\endgroup}

\newc{\etal}{\textrm{et al.}}
\newc{\eg}{\textrm{e.g.}}
\newc{\ie}{\textrm{i.e.}}
\newc{\etc}{\textrm{etc}}
\newc\vs{\textrm{vs.}}
\newc{\cl}{\rm {C.L.}}

\newc{\ev}{\ensuremath{\,\mathrm{eV}}}
\newc{\kev}{\ensuremath{\,\mathrm{keV}}}
\newc{\mev}{\ensuremath{\,\mathrm{MeV}}}
\newc{\gev}{\ensuremath{\,\mathrm{GeV}}}
\newc{\tev}{\ensuremath{\,\mathrm{TeV}}}
\newc{\MeV}{\mev}
\newc{\TeV}{\tev}
\newc{\invpb}{\ensuremath{/\text{pb}}}
\newc{\invfb}{\ensuremath{\,\text{fb}^{-1}}}
\newc\nb{\ensuremath{\,\mathrm{nb}}} \newc\pb{\ensuremath{\,\mathrm{pb}}} \newc\fb{\ensuremath{\,\mathrm{fb}}}
\newc\pc{\ensuremath{\,\mathrm{pc}}}
\newc\kpc{\ensuremath{\,\mathrm{kpc}}}
\newc\mpc{\ensuremath{\,\mathrm{Mpc}}}
\newc\ps{\ensuremath{\,\mathrm{ps}}}
\newc\cmeter{\ensuremath{\,\mathrm{cm}}}
\newc\meter{\ensuremath{\,\mathrm{m}}}
\newc\kmeter{\ensuremath{\,\mathrm{km}}}
\newc\second{\ensuremath{\,\mathrm{s}}}
\newc\msecond{\ensuremath{\,\mathrm{ms}}}
\newc\nsecond{\ensuremath{\,\mathrm{ns}}}
\newc\psecond{\ensuremath{\,\mathrm{ps}}}

\newc{\chisqmin}{\ensuremath{\chi^2_{\mathrm{min}}}}
\newc{\Delchisq}{\ensuremath{\Delta\chi^2}}
\newc{\chisq}{\ensuremath{\chi^2}}
\newc{\like}{\ensuremath{\mathcal{L}}}

\newc\lsim{\ensuremath{\mathrel{\rlap{\lower4pt\hbox{\hskip1pt$\sim$}}\raise1pt\hbox{$<$}}}}
\newc\gsim{\ensuremath{\mathrel{\rlap{\lower4pt\hbox{\hskip1pt$\sim$}}\raise1pt\hbox{$>$}}}}
\newc{\VEV}[1]{\ensuremath{\langle #1 \rangle}}
\newc{\dl}{\ensuremath{\stackrel{\leftarrow}{D}}}
\newc{\dr}{\ensuremath{\stackrel{\rightarrow}{D}}}

\newc{\bcenter}{\begin{center}}    \newc{\ecenter}{\end{center}}
\newc{\bfl}{\begin{flushleft}}    \newc{\efl}{\end{flushleft}}
\newc{\bfr}{\begin{flushright}}    \newc{\efr}{\end{flushright}}

\newc{\bi}{\begin{itemize}}
\newc{\ei}{\end{itemize}}
\newc{\bed}{\begin{description}}
\newc{\eed}{\end{description}}
\newc{\ben}{\begin{enumerate}}
\newc{\een}{\end{enumerate}}

\newc{\be}{\begin{equation}}
\newc{\ee}{\end{equation}}
\newc{\bea}{\begin{eqnarray}}
\newc{\eea}{\end{eqnarray}}
\newc{\ra}{\rightarrow}

\newc{\alphas}{\ensuremath{\alpha_s}}
\newc{\alphatwo}{\ensuremath{\alpha_2}}
\newc{\alphaone}{\ensuremath{\alpha_1}}
\newc{\alphai}[1]{\ensuremath{\alpha_{#1}}}
\newc{\alphaem}{\ensuremath{\alpha_{\mathrm{em}}}}
\newc{\alphaeff}{\ensuremath{\alpha_{\mathrm{eff}}}}
\newc{\sineff}{\ensuremath{\sin \theta_{\mathrm{eff}}}}
\newc{\sinsqeff}{\ensuremath{\sin^2 \theta_{\mathrm{eff}}}}
\newc{\dalphahad}{\ensuremath{\Delta \alpha_{\mathrm{had}}}}
\newc{\yt}{\ensuremath{h_t}} \newc{\yb}{\ensuremath{h_b}} \newc{\ytau}{\ensuremath{h_{\tau}}}
\newc\mz{\ensuremath{M_Z}}
\newc\mw{\ensuremath{m_W}}
\newc\mZ{\mz}        \newc\mW{\mw}
\newc\mhsm{\ensuremath{ m_{H_{\mathrm{SM}}}}}
\newc{\mtop}{\ensuremath{ m_t}}               \newc{\mtpole}{\ensuremath{ M_t}}
\newc{\mbottom}{\ensuremath{ m_b}}
\newc{\mtau}{\ensuremath{ m_{\tau}}}
\newc{\mt}{\mtpole}
\newc{\mb}{\mbottom}
\newc{\rgg}{\ensuremath{R_{h}(\gamma\gamma)}}
\newc{\rzz}{\ensuremath{R_{h}(ZZ)}}
\newc{\rtwogg}{\ensuremath{R_{h_2}(\gamma\gamma)}}
\newc{\rtwozz}{\ensuremath{R_{h_2}(ZZ)}}
\newc{\ronegg}{\ensuremath{R_{h_1}(\gamma\gamma)}}
\newc{\ronezz}{\ensuremath{R_{h_1}(ZZ)}}
\newc{\rsiggg}{\ensuremath{R_{h_\textrm{sig}}(\gamma\gamma)}}
\newc{\rsigzz}{\ensuremath{R_{h_\textrm{sig}}(ZZ)}}
\newc{\llbar}{\ensuremath{\ell\bar{\ell}}}
\newc{\tauptaum}{\ensuremath{ \tau^+\tau^-}}
\newc{\qqbar}{\ensuremath{ q\bar{q}}} \newc{\ppbar}{\ensuremath{ p\bar{p}}}
\newc{\bbbar}{\ensuremath{ b\bar{b}}} \newc{\ttbar}{\ensuremath{ t\bar{t}}}
\newc{\ffbar}{\ensuremath{ f\bar{f}}} \newc{\tautaubar}{\ensuremath{ \tau\bar{\tau}}}
\newc{\mchi}{\ensuremath{m_{\chi}}}
\newc{\squark}{\ensuremath{\tilde{q}}}
\newc{\slepton}{\ensuremath{\tilde{l}}}
\newc{\gluino}{\ensuremath{\tilde{g}}}
\newc{\wino}{\ensuremath{\tilde{W}}}
\newc{\bino}{\ensuremath{\tilde{B}}}
\newc{\mgluino}{\ensuremath{{m_{\gluino}}}}
\newc{\tone}{\ensuremath{{\tilde{t}_1}}}

\newc{\sthw}{\ensuremath{ \sin\theta_W}}              \newc{\cthw}{\ensuremath{\cos\theta_W}}
\newc{\tanthw}{\ensuremath{ \tan\theta_W}}              \newc{\cotthw}{\ensuremath{\cot\theta_W}}
\newc{\ssqthw}{\ensuremath{\sin^2 \theta_W}}
\newc{\msbar}{\ensuremath{\overline{MS}}} \newc{\drbar}{\ensuremath{\overline{DR}}}
\newc{\mtmtsmmsbar}{\ensuremath{ m_t(m_t)^{\msbar}_{{\mathrm{SM}}}}}
\newc{\mtmtsmdrbar}{\ensuremath{ m_t(m_t)^{\drbar}_{{\mathrm{SM}}}}}
\newc{\mtmtmssmdrbar}{\ensuremath{ m_t(m_t)^{\drbar}_{{\mathrm{SUSY}}}}}
\newc{\mbmbmsbar}{\ensuremath{ m_b(m_b)^{\msbar} }}
\newc{\mbmbsmmsbar}{\ensuremath{ m_b(m_b)^{\msbar}_{{\mathrm{SM}}}}}
\newc{\mbmzsmmsbar}{\ensuremath{ m_b(\mz)^{\msbar}_{{\mathrm{SM}}}}}
\newc{\mbmzsmdrbar}{\ensuremath{ m_b(\mz)^{\drbar}_{{\mathrm{SM}}}}}
\newc{\mbmzmssmdrbar}{\ensuremath{ m_b(\mz)^{\drbar}_{{\mathrm{SUSY}}}}}
\newc{\mtaumzsmmsbar}{\ensuremath{ m_{\tau}(\mz)^{\msbar}_{{\mathrm{SM}}}}}
\newc{\mtaumzsmdrbar}{\ensuremath{ m_{\tau}(\mz)^{\drbar}_{{\mathrm{SM}}}}}
\newc{\mtaumzmssmdrbar}{\ensuremath{ m_{\tau}(\mz)^{\drbar}_{{\mathrm{SUSY}}}}}
\newc{\alphasmzms}{\ensuremath{\alpha_s(M_Z)^{\overline{MS}}}}
\newc{\alphaimzms}[1]{\ensuremath{\alpha_{#1}(M_Z)^{\overline{MS}}}}
\newc{\alphaemmz}{\ensuremath{\alpha_{\mathrm{em}}(M_Z)^{\overline{MS}}}}

\newc{\mzero}{\ensuremath{{m_0}}}
\newc{\mhalf}{\ensuremath{ m_{1/2}}}
\newc{\tanb}{\ensuremath{\tan\beta}}
\newc{\azero}{\ensuremath{ A_0}}
\newc{\bzero}{\ensuremath{ B_0}}
\newc{\signmu}{\ensuremath{\rm{sgn}\,\mu}}
\newc{\mueff}{\ensuremath{\mu_{\rm{eff}}}}
\newc{\lam}{\ensuremath{{\lambda}}}
\newc{\kap}{\ensuremath{{\kappa}}}
\newc{\alam}{\ensuremath{{A_{\lambda}}}}
\newc{\akap}{\ensuremath{{A_{\kappa}}}}
\newc{\hs}{\ensuremath{ H_s}}
\newc{\mhs}{\ensuremath{ m_{H_s}}}
\newc{\mgut}{\ensuremath{ M_{\rm GUT}}}
\newc{\mplanck}{\ensuremath{ M_{\rm P}}}      \newc{\mpl}{\ensuremath{ M_{\rm Pl}}}
\newc{\msusy}{\ensuremath{ M_{\rm SUSY}}}      \newc{\ms}{\ensuremath{ M_{\rm S}}}
 \newc{\mhl}{\ensuremath{m_\hl}}
 \newc{\mhone}{\ensuremath{m_{h_1}}}
 \newc{\mhtwo}{\ensuremath{m_{h_2}}}
 \newc{\mglu}{\ensuremath{m_{\tilde g}}}
 \newc{\mul}{\ensuremath{m_{\tilde{u}_L}}}
 \newc{\mtone}{\ensuremath{m_{\tilde{t}_1}}}
 \newc{\ma}{\ensuremath{m_A}}
 \newc{\maone}{\ensuremath{m_{a_1}}}
 \newc{\matwo}{\ensuremath{m_{a_2}}}
 \newc{\hone}{\ensuremath{h_1}}
 \newc{\htwo}{\ensuremath{h_2}}
 \newc{\aone}{\ensuremath{a_1}}
 \newc{\atwo}{\ensuremath{a_2}}
 \newc{\mhu}{\ensuremath{ m_{H_u}}}
 \newc{\mhd}{\ensuremath{ m_{H_d}}}
 \newc{\mhusq}{\ensuremath{ m_{H_u}^2}}
 \newc{\mhdsq}{\ensuremath{ m_{H_d}^2}}
 \newc{\mhuew}{\ensuremath{ m^{\ast}_{H_u}}}
 \newc{\mhdew}{\ensuremath{ m^{\ast}_{H_d}}}
 \newc{\mhuewsq}{\ensuremath{ m^{\ast\, 2}_{H_u}}}
 \newc{\mhdewsq}{\ensuremath{ m^{\ast\, 2}_{H_d}}}
 \newc{\hu}{\ensuremath{ H_u}}
 \newc{\hd}{\ensuremath{ H_d}}
 \newc{\barmhu}{\ensuremath{ \bar{m}_{H_u}}}
 \newc{\barmhd}{\ensuremath{ \bar{m}_{H_d}}}

 \newc{\mqthree}{\ensuremath{m_{\widetilde{Q}_3}^2}}
 \newc{\muthree}{\ensuremath{m_{\tilde{u}_3}^2}}
 \newc{\mdthree}{\ensuremath{m_{\tilde{d}_3}^2}}
 \newc{\mlthree}{\ensuremath{m_{\widetilde{L}_3}^2}}
 \newc{\methree}{\ensuremath{m_{\tilde{e}_3}^2}}
 \newc{\mqtwo}{\ensuremath{m_{\widetilde{Q}_2}^2}}
 \newc{\mutwo}{\ensuremath{m_{\tilde{u}_2}^2}}
 \newc{\mdtwo}{\ensuremath{m_{\tilde{d}_2}^2}}
 \newc{\mltwo}{\ensuremath{m_{\widetilde{L}_2}^2}}
 \newc{\metwo}{\ensuremath{m_{\tilde{e}_2}^2}}
 \newc{\mqone}{\ensuremath{m_{\widetilde{Q}_1}^2}}
 \newc{\muone}{\ensuremath{m_{\tilde{u}_1}^2}}
 \newc{\mdone}{\ensuremath{m_{\tilde{d}_1}^2}}
 \newc{\mlone}{\ensuremath{m_{\widetilde{L}_1}^2}}
 \newc{\meone}{\ensuremath{m_{\tilde{e}_1}^2}}
 \newc{\mone}{\ensuremath{M_1}}
 \newc{\monesq}{\ensuremath{M_1^2}}
 \newc{\mtwo}{\ensuremath{M_2}}
 \newc{\mtwosq}{\ensuremath{M_2^2}}
 \newc{\mthree}{\ensuremath{M_3}}
 \newc{\mthreesq}{\ensuremath{M_3^2}}
 \newc{\atau}{\ensuremath{{A_{\tau}}}}
 \newc{\at}{\ensuremath{{A_{t}}}}
 \newc{\ab}{\ensuremath{{A_{b}}}}
 \newc{\atausq}{\ensuremath{{A_{\tau}^2}}}
 \newc{\atsq}{\ensuremath{{A_{t}^2}}}
 \newc{\absq}{\ensuremath{{A_{b}^2}}}

 \newc{\dmzero}{\ensuremath{\Delta{_{m_0}}}}
 \newc{\dmhalf}{\ensuremath{\Delta{_{m_{1/2}}}}}
 \newc{\dmu}{\ensuremath{\Delta{_{\mu}}}}

 \newc{\pten}{\ensuremath{\psi_{10}}}
 \newc{\ffive}{\ensuremath{\phi_{5}}}
 \newc{\hfive}{\ensuremath{h_{5}}}
 \newc{\hbfive}{\ensuremath{h_{\bar{5}}}}
 \newc{\thet}{\ensuremath{\theta_{50}}}
 \newc{\thetb}{\ensuremath{\theta_{\,\overline{50}}}}
 \newc{\ptenhat}{\ensuremath{\hat{\psi}_{10}}}
 \newc{\ffivehat}{\ensuremath{\hat{\phi}_{5}}}
 \newc{\hfivehat}{\ensuremath{\hat{h}_{5}}}
 \newc{\hbfivehat}{\ensuremath{\hat{h}_{\bar{5}}}}
 \newc{\thethat}{\ensuremath{\hat{\theta}_{50}}}
 \newc{\thetbhat}{\ensuremath{\hat{\theta}_{\,\overline{50}}}}
 \newc{\si}{\ensuremath{\Sigma}}
 \newc{\mfive}{\ensuremath{m_5^2}}
 \newc{\mten}{\ensuremath{m_{10}^2}}
 \newc{\dfive}{\ensuremath{\Delta^2_5}}
 \newc{\dbfive}{\ensuremath{\Delta^2_{\bar{5}}}}
 \newc{\dfifty}{\ensuremath{\Delta^2_{50}}}
 \newc{\dfiftyb}{\ensuremath{\Delta^2_{\,\overline{50}}}}
 \newc{\msi}{\ensuremath{m_{\Sigma}^2}}
 \newc{\lamh}{\ensuremath{\lambda_{H}}}
 \newc{\lamhb}{\ensuremath{\lambda_{\bar{H}}}}
 \newc{\ah}{\ensuremath{A_{H}}}
 \newc{\ahb}{\ensuremath{A_{\bar{H}}}}
 \newc{\lams}{\ensuremath{\lambda_{S}}}
 \newc{\as}{\ensuremath{A_{S}}}
 \newc{\lamsig}{\ensuremath{\lambda_{\si}}}
 \newc{\asig}{\ensuremath{A_{\si}}}

 \newc{\msten}{\ensuremath{m_{16}^2}}
 \newc{\mhun}{\ensuremath{m_{126}^2}}
 \newc{\mhunb}{\ensuremath{m_{\bar{126}}^2}}
 \newc{\mthun}{\ensuremath{m_{210}^2}}
 \newc{\ahun}{\ensuremath{A_{\bar{126}}}}
 \newc{\yhun}{\ensuremath{Y_{\bar{126}}}}
 \newc{\aten}{\ensuremath{A_{10}}}
 \newc{\yten}{\ensuremath{Y_{10}}}
 \newc{\alone}{\ensuremath{A_{\lambda_1}}}
 \newc{\altwo}{\ensuremath{A_{\lambda_2}}}
 \newc{\althree}{\ensuremath{A_{\lambda_3}}}
 \newc{\althreeb}{\ensuremath{A_{\bar{\lambda_3}}}}
 \newc{\lone}{\ensuremath{\lambda_1}}
 \newc{\ltwo}{\ensuremath{\lambda_2}}
 \newc{\lthree}{\ensuremath{\lambda_3}}
 \newc{\lthreeb}{\ensuremath{\bar{\lambda_3}}}

\newc{\sigsip}{\ensuremath{\sigma^{\rm SI}_{p}}}	\newc{\sigsin}{\ensuremath{\sigma^{\rm SI}_{n}}}
\newc{\sigsdp}{\ensuremath{\sigma^{\rm SD}_{p}}}	\newc{\sigsdn}{\ensuremath{\sigma^{\rm SD}_{n}}}
\newc{\sigsi}{\ensuremath{\sigma^{\rm SI}}}	\newc{\sigsd}{\ensuremath{\sigma^{\rm SD}}}
\newc{\sigv}{\ensuremath{\sigma v}}
\newc{\abund}{\ensuremath{ \Omega h^2}}
\newc{\omegadm}{\ensuremath{ \Omega_{{\rm DM}}}}     \newc{\abunddm}{\ensuremath{ \Omega_{{\rm DM}} h^2}}
\newc{\omegam}{\ensuremath{ \Omega_{{\rm m}}}}       \newc{\abundm}{\ensuremath{ \Omega_{{\rm m}} h^2}}
\newc{\omegab}{\ensuremath{ \Omega_{{\rm b}}}}	\newc{\abundb}{\ensuremath{ \Omega_{{\rm b}} h^2}}
\newc{\omegatot}{\ensuremath{ \Omega_{{\rm TOT}}}}
\newc{\omegacdm}{\ensuremath{ \Omega_{{\rm CDM}}}}   \newc{\abundcdm}{\ensuremath{ \Omega_{{\rm CDM}} h^2}}
\newc{\omegalambda}{\ensuremath{ \Omega_{\Lambda}}} \newc{\abundlambda}{\ensuremath{ \Omega_{\Lambda} h^2}}
\newc{\omegarad}{\ensuremath{ \Omega_{{\rm rad}}}}  \newc{\abundrad}{\ensuremath{ \Omega_{{\rm rad}} h^2}}
\newc{\rhocrit}{\ensuremath{ \rho_{\rm crit}}}
\newc{\rhochi}{\ensuremath{ \rho_{\chi}}}
\newc{\abunchi}{\ensuremath{\Omega_\chi h^2}}
\newc{\abundlsp}{\ensuremath{\Omega_{\rm LSP}h^2}}
\newc{\abundchi}{\ensuremath{\Omega_\chi h^2}}
\newc{\tf}{\ensuremath{T_f}} \newc{\xf}{\ensuremath{x_f}}
\newc{\tr}{\ensuremath{T_R}}

\newc{\amu}{\ensuremath{ a_{\mu}}}        \newc{\amususy}{\ensuremath{ a_{\mu}^{\mathrm{SUSY}}}}
\newc{\amuexpt}{\ensuremath{ a_{\mu}^{\mathrm{expt}}}}        \newc{\amusm}{\ensuremath{ a_{\mu}^{\mathrm{SM}}}}
\newc\deltaamu{\ensuremath{\Delta a_{\mu}}} \newc{\deltaamususy}{\ensuremath{\delta a_{\mu}^{\mathrm{SUSY}}}}
\newc\gmtwo{\ensuremath{ (g-2)_{\mu}}}
\newc{\deltagmtwomususy}{\ensuremath{\delta\left(g-2\right)_{\mu}^{\mathrm{SUSY}}}}
\newc{\deltagmtwomu}{\ensuremath{\delta\left(g-2\right)_{\mu}}}
\newc\BR{\ensuremath{\rm BR}}
\newc\bsgamma{\ensuremath{ b\rightarrow s \gamma }}
\newc\bxsgamma{\ensuremath{\overline{B}\rightarrow X_{s}\gamma}}
\newc\brbsgamma{\ensuremath{\BR\left(\bsgamma\right)}}
\newc\brbxsgamma{\ensuremath{\BR\left(\bxsgamma\right)}}
\newc\bsmumu{\ensuremath{B_s\to\mu^+\mu^-}}
\newc\brbsmumu{\ensuremath{\BR\left(B_s\to\mu^+\mu^-\right)}}
\newc\bdmmumu{\ensuremath{\overline{B}_d\to\mu^+\mu^-}}
\newc\bbbarmix{\ensuremath{\overline{B}_s\mbox{-}B_s}}      
\newc\delmbs{\ensuremath{\Delta M_{B_s}}}
\newc{\butaunu}{\ensuremath{B_u \rightarrow \tau \nu}}
\newc{\brbutaunu}{\ensuremath{\BR\left(B_u \rightarrow \tau \nu\right)}}

\let\oldcite\cite
\renewcommand*{\cite}{~\oldcite}

\newcommand*{\hl}{\ensuremath{h}}

\graphicspath{{figures/}}

\definecolor{ferrarired}{rgb}{1.0, 0.11, 0.0}

\title{Neutrino beam-dump experiment with FASER at the LHC}

\author[a]{Krzysztof~Jod\l{}owski,}
\author[b,c,a]{Sebastian Trojanowski}

\affiliation[a]{National Centre for Nuclear Research,\\
  Pasteura 7, 02-093 Warsaw, Poland}
\affiliation[b]{Astrocent, Nicolaus Copernicus Astronomical Center Polish Academy of Sciences, ul. Bartycka 18, 00-716 Warsaw, Poland}
\affiliation[c]{Consortium for Fundamental Physics, School of Mathematics and Statistics, University of Sheffield,\\Hounsfield Road, Sheffield, S3 7RH, UK}

\emailAdd{krzysztof.jodlowski@ncbj.gov.pl}
\emailAdd{strojanowski@camk.edu.pl}

\abstract{The neutrino physics program at the LHC, which will soon be initiated by the FASER experiment, will provide unique opportunities for precision studies of neutrino interaction vertices at high energies. This will also open up the possibility to search for beyond the standard model (BSM) particles produced in such interactions in the specific high-energy neutrino beam-dump experiment. In this study, we illustrate the prospects for such searches in models with the dipole or $Z^\prime$ portal to $\gev$-scale heavy neutral leptons. To this end, we employ both the standard signature of new physics that consists of a pair of oppositely-charged tracks appearing in the decay vessel, and the additional types of searches. These include high-energy photons and single scattered electrons. We show that such a variety of experimental signatures could significantly extend the sensitivity reach of the future multi-purpose FASER 2 detector during the High-Luminosity phase of the LHC.
}

\begin{document}
\maketitle
\flushbottom

\section{Introduction}

Despite being so difficult to probe experimentally, interactions of neutrinos have always been a window to new exciting phenomena with potentially groundbreaking consequences for our understanding of physics. Only in recent years, such studies have led to the establishment of neutrino oscillations, first at low energy\cite{Fukuda:1998mi,Ahmad:2001an,Ahmad:2002jz}, but then also for energies up to $\mathcal{O}(100~\gev)$\cite{Aartsen:2013jza}. Instead, at much higher energies, $E_\nu\sim\textrm{PeV}$, the observation of the astrophysical neutrino flux\cite{Aartsen:2013jdh} has already proven the discovery potential of a new type of multi-messenger astronomy\cite{IceCube:2018dnn}.

In the intermediate energy range, $E_\nu\sim\tev$, the properties of neutrinos are generally expected to follow the Standard Model (SM) predictions. However, in fact, only very limited experimental evidence for this is currently available. In particular, neutrino scatterings have been extensively studied for energies up to $E_\nu\sim 300~\gev$ (see Ref.\cite{Zyla:2020zbs} for review). For $\tev$ energies, instead, the relevant cross section measurements employing large-scale neutrino telescopes suffer from substantial uncertainties\cite{Aartsen:2017kpd,Bustamante:2017xuy}, while detailed characteristics of these events are even more challenging to probe.

The recently approved FASER$\nu$ experiment\cite{Abreu:2019yak,Abreu:2020ddv} at the Large Hadron Collider (LHC) aims at starting to fill in this gap during the LHC Run 3. Both the experiment and its potential extension towards the High-Luminosity LHC phase (HL-LHC), dubbed FASER$\nu$2, will benefit from a high-intensity and high-energy beam of forward-going neutrinos produced in $pp$ interactions at the LHC. The mean expected energy of these neutrinos interacting in the distant detector is of order several hundred $\gev$\cite{Abreu:2019yak}.

In this study, we analyze the prospects of probing beyond the Standard Model (BSM) physics in such neutrino interactions. In particular, our focus is on a possible \textsl{secondary production} of new physics species in neutrino scatterings in the close vicinity or inside the FASER$\nu$ detector, as well as on subsequent detectable signals that the new particles could generate. Similar such processes have already been shown to extend the discovery potential of FASER with regards to BSM scenarios predicting a light dark photon, dark matter, and dark Higgs boson\cite{Jodlowski:2019ycu}. We illustrate how a combination of excellent properties of both the FASER$\nu$ and FASER\cite{Ariga:2018zuc,Ariga:2018pin} detectors, with the latter to be placed right behind FASER$\nu$ to search for light and long-lived particles (LLPs)\cite{Feng:2017uoz,Ariga:2018uku}, can improve the expected sensitivity reach and lead to interesting experimental signatures in selected popular models predicting $\gev$-scale new particles coupled to neutrinos.\footnote{At the last stage of the work on this project, a similar study has appeared which concerns the new physics particle production in neutrino interactions inside or in front of FASER$\nu$, cf. Ref.\cite{Bakhti:2020szu}. In that study, the primary focus is on the multi-muon signatures, including through-going muons, and the production of the BSM species in deep inelastic scattering (DIS) processes. Instead, we discuss experimental signatures based on the detection of a photon or an $e^+e^-$ pair, as well as on scatterings off electrons. As far as the secondary production is concerned, we focus on coherent and elastic incoherent processes that minimize the risk of activating detector veto layers in our analysis.}

In fact, the aforementioned discovery of neutrino oscillations remains among the most important hints towards the extension of the SM. In the minimal such extension, the existence of new right-handed neutrinos is postulated\cite{Minkowski:1977sc,GellMann:1980vs,Mohapatra:1979ia,Yanagida:1980xy,Schechter:1980gr}. This leads to an interesting phenomenology and can also help to understand the nature of dark matter and the origin of the baryon asymmetry in the Universe, cf. Ref.\cite{Drewes:2013gca} for review.  In this study, instead, we will focus on BSM scenarios predicting neutrino upscatterings into BSM \textsl{heavy neutral leptons} (HNLs) that are mediated by light vector particles or photons in the dipole portal model (see, e.g., Refs\cite{Gninenko:2009ks,Aparici:2009fh,Gninenko:2010pr,Coloma:2017ppo,Magill:2018jla,Ballett:2018ynz,Bertuzzo:2018itn,Shoemaker:2018vii,Fischer:2019fbw,Datta:2020auq,Dutta:2020scq,Abdallah:2020biq,Abdullahi:2020nyr,Shoemaker:2020kji,Brdar:2020quo,Jho:2020jfz,Plestid:2020vqf} for further discussion about the models and their phenomenology). As we show, in such scenarios the BSM neutrino scatterings can lead to new types of signatures in the FASER and FASER$\nu$ detectors.

This paper is organized as follows. In \cref{sec:modeling}, we provide basic detector details and discuss the experimental signatures of our interest. \Cref{sec:models} is devoted to a discussion of the new physics models that we focus on in this paper. We present the results of our analysis in \cref{sec:results} and conclude in \cref{sec:conclusions}.  The relevant expressions for the production cross sections and decay widths are given in \cref{app:sigma,app:decay}, respectively.

\section{New physics from neutrino scatterings in FASER$\nu$\label{sec:modeling}}

\subsection{General physics motivation -- neutrino interactions with light mediators}

Despite a large number of forward-going neutrinos at the LHC, only few of them will interact in the detector due to their tiny scattering cross section. In particular, during LHC Run 3 about $10^4$ neutrino-nucleus scattering events are expected in FASER$\nu$\cite{Abreu:2019yak}, while this number could grow to $10^5-10^6$ in the HL-LHC phase. Therefore, there is \textsl{a priori} little room for BSM particles to be produced at detectable rates in these processes, especially given their suppressed interaction strength with the SM species characterized by the coupling constant $g_D\ll 1$.

The events rates, however, become significantly larger in the presence of light mediator particle $X$ between neutrinos and other SM fermions. In this case, for low momentum exchange and $g_D^4/m_X^4\sim G_F^2$, where $G_F\simeq 1.166\times 10^{-5}~\gev^{-2}$ is the Fermi coupling constant\cite{Zyla:2020zbs}, the Fermi contact interaction of neutrinos in the SM can be supplemented by a similar BSM contribution to the total cross section. Even if the BSM-induced scattering rates are smaller than the SM ones, new physics effects could still be successfully searched for by employing specific features of BSM interactions. In particular, in the following, we will make use of such signatures relying on displaced decays of neutrino-induced BSM particles or on differences in the scattering kinematics that could be captured with the use of appropriate cuts.

The high-energy of neutrinos produced at the LHC, along with their high beam intensity, allows one to further improve the sensitivity reach in these models, especially for increasing mass of the produced BSM species. In particular, for $\tev$-energy neutrinos scatterings off electrons, one obtains the center-of-mass energy in the collision of order $\sqrt{s} = \sqrt{2E_\nu m_e}\sim \gev$, while it grows even larger for such high-energy neutrino interactions with nuclei. For the growing mass of the BSM particles and their decreasing lifetime, larger boost factors also allow for better reconstruction of their displaced decay signatures.

A powerful way to constrain these scenarios is by searching for increased neutrino scattering cross section off electrons\cite{Lindner:2018kjo} or for a similar effect in coherent scatterings off nuclei\cite{Lindner:2016wff}. Other relevant bounds come from i.a. neutrino experiments, as well as past and present beam-dump or missing-energy searches (cf. Refs\cite{Beacham:2019nyx,Alimena:2019zri,Agrawal:2021dbo} for review of this rich experimental program). These could be associated with both the primary production of new physics species in the target material and with the secondary interactions of neutrinos. Below, we refer to specific such searches relevant for the models under study.

\subsection{Detectors}

The FASER$\nu$ detector will be placed in a side tunnel of the LHC about $480$~m away from the ATLAS interaction point (IP). It will be sensitive to signatures of high-energy neutrino interactions happening in a $1$~m long emulsion cloud chamber (ECC) detector consisting of tungsten layers interleaved with emulsion films, cf. Ref.~\cite{Ariga:2020lbq} for recent review about nuclear emulsions. The detailed design of the neutrino detector and the main FASER spectrometer can be found in Refs\cite{Ariga:2018pin,Abreu:2020ddv}.

The simplified possible design of the future FASER 2 experiment, to take data during HL-LHC and search for LLP decays, which we use, follows Ref.\cite{Ariga:2018uku}. In the case of the potential successor of the neutrino subdetector, which we refer to as FASER$\nu$2, we employ a simplified design from Ref.\cite{Batell:2021blf}, in which the prospects for searching for light dark matter in the far-forward region of the LHC have been analyzed. The assumed design will employ a larger ECC detector than FASER$\nu$ that will be placed along the beam collision axis in a similar distance to the ATLAS IP. The transverse size of the tungsten active material in FASER$\nu$2 is set to be $50~\textrm{cm}\times 50~\textrm{cm}$ and its length is equal to $2$~m. In the case of some of the signatures discussed below, it is also useful to consider a sweeping magnet to be installed along the beam collision axis in front of the detector. Similarly, the emulsion films could be interleaved with layers of the electronic detectors. This helps to suppress the muon-induced backgrounds (BG). We will refer to these design improvements when discussing possible sources of BG in the searches described in our study.

Importantly, alternative detector designs have also been discussed in  Ref.\cite{Batell:2021blf} that would employ liquid argon (LAr) time-projection chamber (TPC) technology. While in our estimates below we focus on the aforementioned ECC detector, we note that similar such searches could be performed in the LAr detector. In the latter case, depending on the particular experimental signature of interest, lower target material density will partially limit the expected sensitivity reach due to secondary particle production inside the detector. On the other hand, improved BG rejection capabilities could make it easier to study e.g. electron scattering signatures.

\subsection{Neutrino flux and the production of BSM particles\label{sec:fluxandprod}}

\paragraph{Neutrino flux} The flux and spectra of high-energy neutrinos going in the far-forward region of the LHC have been simulated by the FASER collaboration\cite{Abreu:2019yak} and by the CERN STI group\cite{Beni:2020yfy}. In our study, we employ the former results when presenting the sensitivity reach plots, although we note that uncertainties in modeling of the neutrino spectrum at the FASER location do not typically have a significant impact on our results. For simplicity, when analyzing the future sensitivity reach of FASER$\nu$2 to operate during the HL-LHC phase, we rescale the aforementioned flux of forward-going neutrinos by the appropriate luminosity factor and a larger transverse size of the detector, while we assume the same energy spectrum.

\paragraph{Primary and secondary production of new physics particles} The BSM particles with order $\gev$ masses can abundantly be produced at the LHC, especially in its forward direction, provided that their relevant coupling constant to the SM, $g_D$, is not suppressed too much. The larger $g_D$ remains, however, the smaller is the lifetime of unstable such new species. This often makes them decaying too early, i.e. before they can travel the entire distance between the ATLAS IP and FASER, unless they are produced in secondary production processes in the close vicinity of the detector. When studying such processes, we estimate the sensitivity reach with dedicated Monte Carlo (MC) simulations and follow the discussion in Ref.\cite{Jodlowski:2019ycu} for the cuts used in the analysis.

In this study, we primarily focus on the BSM particle secondary production in neutrino interactions. Such neutrino interactions in the tungsten layers of FASER$\nu$ can result in a flux of new and boosted heavier species produced right in front of the FASER spectrometer. These will travel $\mathcal{O}(1~\textrm{m})$ distance and decay inside FASER generating visible signal events. Notably, the largest such contribution in models of our interest is associated with coherent neutrino scatterings off nuclei that is characterized by a small momentum exchange, typically $|Q^2|<(100~\mev)^2$. Instead, incoherent contributions are often subdominant. In particular, in currently unconstrained regions of the parameter space of the models of our interest that can be probed by FASER$\nu$2, they contribute with typically $\lesssim 1\%$ of additional signal events obtained due to the secondary HNL production. However, we do take them into account, since they become important in the models with the massive vector mediator, especially when the mediator or the HNL mass grows above $\gev$. We conservatively neglect the DIS contributions as they often generate additional visible activity inside or in front of the detector, which makes it more challenging to avoid vetoing such events.

For completeness, we also take into account possible primary production processes of BSM particles of our interest at the ATLAS IP. To be detectable at FASER, the new physics species produced this way have to be characterized by a sufficiently large lifetime, such that they can survive without decaying until they reach the detector. When modeling the primary production of LLPs, we obtain the spectrum of light mesons with the \texttt{CRMC} simulation package\cite{CRMC} and the MC event generator \texttt{EPOS-LHC}\cite{Pierog:2013ria}. We follow the discussion in Refs\cite{Blumlein:2013cua,Feng:2017uoz,deNiverville:2016rqh} for the Fermi-Weizsacker-Williams (FWW) approximation when treating the production of light vector particles in proton bremsstrahlung.

\subsection{Experimental signatures of new physics}

\subsubsection{LLP signal inside the FASER decay vessel -- $e^+e^-$ and $\gamma$}

\paragraph{Signal} A vanilla LLP decay signature to be searched for in FASER consists of a pair of high-energy and oppositely-charged tracks from the LLP decays, e.g. an $e^+e^-$ pair. These are detected in the spectrometer and could also deposit energy in the calorimeter. FASER will also be sensitive to photon-induced signals from LLP decays, cf. a similar recent discussion in Ref.\cite{Kling:2020mch}. In our simulations, we require $E_{\textrm{vis}}>100~\gev$ threshold for the energy of the visible particles produced in the decay vessel.

\paragraph{Backgrounds} Notably, for such high energies, the search for a pair of high-energy oppositely charged tracks can be performed with negligible BG and, therefore, remains a very sensitive probe of new physics\cite{Ariga:2018zuc,Ariga:2018pin}. Instead, the neutrino-induced BG in the search for single high-energy photons appearing in the decay vessel can be minimized by the use of a dedicated preshower detector. This would allow to reject relevant BG events from e.g. electron neutrino charged current (CC) scatterings happening sufficiently deep inside the calorimeter, with no detectable backsplash effect. The possible BG from muon-induced photons are expected to be vetoed with a very high efficiency by detecting a time-coincident muon going through the detector\cite{Ariga:2018pin}. Hence, the excess of even such single-photon events unaccompanied by any muon, will already be indicative of new physics, especially if they correspond to only a limited visible energy range. It remains difficult to reliably estimate the number of such possible residue BG events without a full detector simulation that takes into account the presence of the preshower detector, which would go beyond the scope of this study. Instead, in the following, when presenting the results for this signature, we will show the lines with a fixed expected number of such events in the detector and discuss their energy spectrum.

\subsubsection{Prompt decays of high-energy LLPs inside the ECC detector}

\paragraph{Signal} After the secondary production of LLPs, their prompt decays can also happen inside the ECC detector. In the following, we consider a search for such a signature consisting of very high-energy photons with $E_\gamma>1~\tev$, or even $E_\gamma>3~\tev$, that are unaccompanied by any time-coincident muon. To this end, we focus on the FASER$\nu$2 experiment during the HL-LHC phase.

The relevant interaction vertex inside the emulsion detector corresponds to a single photon with no hadronic activity. The latter condition can be satisfied by requiring a low momentum transfer in the scattering process, $|Q|^2<(100~\mev)^2$. Such a signature has already been discussed in Ref.\cite{Magill:2018jla} in the context of the proposed SND@SHiP detector\cite{Ahdida:2020evc} and the neutrino dipole portal model to HNLs, cf. \cref{sec:dipoleportal}. Since it allows one to study even immediate decays of the LLPs, it extends the sensitivity reach in new physics scenarios towards the larger values of the LLP masses and the increasing value of their coupling constant to the SM.

\paragraph{Backgrounds} Since employing the single-photon signature in FASER$\nu$2 has not been discussed yet in the literature, we comment below on the possible BG for this search and on the methods that could be used to mitigate their impact. We stress that the ECC detectors only collect data integrated over time. Hence, inside them, BG from muon-induced photons can easily mimic the BSM signal of our interest, unless the muons can be actively vetoed and partially swept away, as we have already mentioned above.

Additional handles over the muon-induced BG can be obtained by increasing the energy threshold of photon-initiated EM showers relevant for the new physics search. In particular, while the total number of the through-going muons during the HL-LHC phase can be as large as $N_{\mu,\textrm{HL-LHC}}\sim 10^{11}$, it is suppressed by one or three orders of magnitude for $E_\mu>\tev$ or $3~\tev$, respectively. By following the simulation setup discussed in Ref.\cite{Batell:2021blf} and by employing the FLUKA code\cite{Ferrari:2005zk,Battistoni:2015epi}, we have estimated the production rate of muon-induced high-energy photons with $E_\gamma>1~\tev$ ($3~\tev$) inside FASER$\nu$2 during HL-LHC to be $10^{-3}$ ($10^{-7}$) per a single through-going muon track in the detector. Here, we have used a properly rescaled high-energy spectrum of such muons, which was simulated for the FASER detector and for LHC Run 3\cite{FLUKAstudy,Ariga:2018pin}. The remaining muon-induced high-energy photons in the detector could be further rejected as BG by detecting the time-coincident parent muon, cf. Ref.\cite{Batell:2021blf} for further discussion. We assume that this is the case in the following. Instead, the corresponding search in FASER$\nu$ during LHC Run 3 will be much more challenging due to its smaller size and the relevant integrated luminosity, as well as due to the lack of the muon veto capabilities of the emulsion detector with no interleaved electronic tracker layers.

The neutrino scattering events could also constitute BG for the single-photon search. As far as very high energy thresholds are concerned, the neutrino interactions are dominated by the DIS events. In particular, in the $\nu_e$ CCDIS scatterings, a high-energy outgoing electron is produced. The electron, however, will generally not mimic a photon-initiated shower in FASER$\nu$2 thanks to the excellent capabilities of the ECC detector to reconstruct both the electron and positron tracks from the initial pair-production process $\gamma N\to e^+e^- N$. We note, however, that such capabilities might be reduced for the very high-energy and collimated electron-positron pair. Even in this case, however, in the high-energy DIS processes for $E_\nu\sim\textrm{a few}\tev$, typically a large momentum transfer to the nucleus takes place. This generates additional visible hadronic activity emerging from the vertex, e.g. charged pions. These charged tracks penetrate the detector and could be used for the vetoing purposes by employing the interleaved electronic detectors. Instead, in the BSM signal events that we study below, only a small momentum transfer is favored, as dictated by the low mediator mass.

We have also estimated using \texttt{GENIE}\cite{Andreopoulos:2009rq,Andreopoulos:2015wxa} the residue BG for such events from the subdominant quasi-elastic scatterings of the high-energy electron neutrinos. We expect a few tens of such events with only a single electron or positron with $E_e>\tev$ detectable at the vertex, while this number drops down to below $10$ events for the energy threshold of $3~\tev$. These could mimic the BSM signal events if the outgoing $e^\pm$ is reconstructed as a photon.

While a detailed detector-level simulation of this signature would be needed to design the analysis cuts that maximize the expected sensitivity reach, below, for illustration, we present results by showing the lines with the constant number of the expected BSM signal events. These correspond to the aforementioned simple cuts on the photon energy of $E_\gamma>1~\tev$ or $3~\tev$. Notably, as discussed in \cref{sec:results} for the models of our interest, the number of the signal events can reach even up to $10^4$ BSM-induced high-energy photons in the detector.

\subsubsection{Scattering off electrons}

\paragraph{Signal} While LLP decays into charged tracks or photons constitute the main signature of our interest, we also discuss the additional sensitivity reach that can be associated with new-physics-induced neutrino scatterings off electrons producing detectable electron recoils inside the neutrino detector. Here, we follow the discussion in Ref.\cite{Batell:2021blf} when defining our cuts on the final-state electron recoil energy and angle.

As discussed in \cref{sec:results}, the scattering signature could also appear as a result of interactions of new unstable species, provided that they have significantly large lifetime to typically reach the detector without first decaying. It will then closely resemble the DM scattering signature and will add to the total expected neutrino-like event rate. For the most abundant, energetic and collimated such BSM events, it could also be beneficial to perform the search by releasing the cuts on the electron recoil angle, which allows the new physics events to more closely resemble the neutrino-induced BG.

\paragraph{Backgrounds} A detailed BG discussion for this experimental signature can be found in Ref.\cite{Batell:2021blf}. Here, we briefly recapitulate the main results of that analysis. In particular, before the angular cuts are imposed on the recoiled electron, one expects few tens of neutrino-induced BG events during HL-LHC that are characterized with no additional detectable charged tracks emerging from the vertex beside the electron with the recoil energy in between $0.3$ and $20~\gev$. This number drops down to the level of $\mathcal{O}(10)$ after additional angular cuts are taken into account. Below, when referring to this discussion, we will assume that the BG event rate is dominated by the statistical fluctuation. For simplicity, we will then consider the bounds on new physics set based on about $20$ additional expected BSM events.

\medskip

Having discussed the experimental signatures of our interest, we now present the new physics scenarios that we use to illustrate the neutrino beam-dump physics program at FASER.

\section{Selected beyond the Standard Model scenarios\label{sec:models}}

As discussed above, in order to take advantage of the new physics production in neutrino scatterings in FASER$\nu$, we will focus below on BSM models predicting new $\gev$-scale HNLs that can be produced in upscatterings of the SM neutrinos mediated by a light vector particle $X$.\footnote{We note that, even in the absence of light mediator fields, the secondary production of heavy fermions in front of FASER$\nu$ could take place due to the mixing with the active neutrinos induced by the neutrino coupling portal, $NHL$. However, given current bounds on the respective mixing angles of $\gev$-scale such HNLs and due to the limited event statistics, we expect this contribution to play a negligible role in our considerations. The FASER sensitivity reach for such HNLs coming from the primary production at the ATLAS IP have been discussed in Refs\cite{Kling:2018wct,Helo:2018qej,Ariga:2018uku}.} To this end, we will assume that $X$ is either the massless SM photon or a new light dark gauge boson $A^\prime$. The choice of the models below is also dictated by distinct phenomenological aspects of FASER searches that can be discussed in connection to them. We note, however, that scalar mediators could also lead to interesting phenomenology, see e.g. Ref.\cite{Farzan:2018gtr} for recent such study.

\subsection{Neutrino dipole couplings to heavy neutral leptons\label{sec:dipoleportal}}

One of the consequences of extending the SM with additional right-handed neutrinos, is that the neutrino magnetic moment is generated with a tiny value proportional to the neutrino mass\cite{Petcov:1976ff,Fujikawa:1980yx,Pal:1981rm,Shrock:1982sc,Dvornikov:2003js}. This value can grow even larger, up to a detectable level, in more complex BSM models\cite{Giunti:2014ixa,Lindner:2017uvt,Babu:2020ivd,Brdar:2020quo}. Recently, such scenarios predicting HNLs with the dipole coupling to SM neutrinos have received renewed attention and have been studied in the context of beam-dump and neutrino experiments, astrophysics, cosmology, and direct searches at dark matter experiments\cite{Coloma:2017ppo,Magill:2018jla,Shoemaker:2018vii,Shoemaker:2020kji,Brdar:2020quo}.

At the effective low-energy level, neutrino dipole portal to HNLs is described by the following Lagrangian
\begin{equation}
\mathcal{L} \supset \mu_N\,\bar{\nu}_L \sigma_{\mu \nu}N_R F^{\mu \nu}+\textrm{h.c.},
\label{eq:Lagdipole}
\end{equation}
where $\sigma^{\mu \nu}=\frac i2 [\gamma^\mu, \gamma^\nu]$, $F^{\mu \nu}$ is electromagnetic field strength tensor, $\nu$ is the SM neutrino and $N_R$ represents the sterile neutrino, which is SM gauge singlet. The coupling strength $\mu_N$ has units of $\textrm{mass}^{-1}$ and is bounded to be $\mu_N\lesssim 10^{-6}\gev^{-1}$ for $\gev$-scale HNLs\cite{Magill:2018jla,Shoemaker:2018vii,Brdar:2020quo}. While the UV-completion scale for this model can be as large as $1/\mu$, already at electroweak energies the dipole interaction in \cref{eq:Lagdipole} should be promoted to dimension-6 operator by the Higgs insertion such that the SM gauge invariance is restored. In the following, we consider the simplified model given by \cref{eq:Lagdipole} even for the interaction of neutrinos with $E_\nu\sim\tev$, since the typical momentum exchange of our interest is much smaller than the electroweak scale.

The dipole interaction of \cref{eq:Lagdipole} leads to inelastic upscattering transitions of the active neutrinos to the HNLs, $\nu Z\to N Z$, where $Z$ is a target nucleus.\footnote{Hereafter, we refer to the coherent superposition of the light neutrinos as the active neutrino in a definite flavor state. We also note that neutrino oscillations play a negligible role in our analysis given the distance $L\sim 0.5~\kmeter$ between the production and interaction point, and typical energy $E_\nu$ of order a few hundred GeV.} Once produced, boosted HNLs travel at a finite distance and subsequently decay inside the detector into a single photon and neutrino, $N\to\gamma\nu$, with the typical decay length given by
\begin{equation}
\bar{d}_{N,\textrm{dipole}} \simeq (1~\textrm{m})\,\left(\frac{E_N}{500~\gev}\right)\,\left(\frac{1~\gev}{m_N}\right)^4\,\left(\frac{10^{-6}~\gev^{-1}}{\mu_N}\right)^2.
\label{eq:Ndecay2body}
\end{equation}
For the given values of the parameters of the model, typical energy of the decaying HNL is dictated by an interplay between the energy-dependent $\bar{d}_{N,\textrm{dipole}}(E_N)$ and the distance $l\sim 1~\textrm{m}$ between the the tungsten plates of FASER$\nu$, in which the upscattering can most efficiently occur, and the FASER decay vessel. The condition $\bar{d}_{N,\textrm{dipole}}(E_N)\sim l$ translates into a limited energy range of visible photons seen in the detector, as we illustrate in \cref{sec:nutolight} (cf. also \cref{sec:bimodal} for a similar discussion regarding the bi-modal $e^+e^-$ spectrum in the model with the dark vector portal).

In addition, three-body decays of HNLs, e.g. $N\to\nu\ell\ell$, are also possible that could generate the signal consisting of two oppositely-charged tracks, although with a suppressed branching fraction, $\mathcal{B}(N\to\nu\ell\ell)\sim 10^{-3}$ to $10^{-2}$. We provide the relevant expressions for the scattering cross sections and decay widths in this model in \cref{app:sigma,app:decay}.

The transition magnetic moment between the active and sterile neutrinos have been considered\cite{Gninenko:2009ks,Gninenko:2010pr} in connection to a possible BSM explanation of the MiniBooNE\cite{AguilarArevalo:2007it} and LSND\cite{Athanassopoulos:1996jb} anomalies. Interestingly, the former has recently been strengthened to the $4.8\sigma$ level after a release of the updated analysis by the MiniBooNE collaboration\cite{Aguilar-Arevalo:2020nvw}. The relevant study for the pure dipole-portal has been performed in Ref.\cite{Magill:2018jla}, in which the simultaneous explanation of both anomalies in this scenario has been excluded. Instead, the MiniBooNE-only region of interest (RoI) corresponding to the HNL with $m_N\simeq 500~\mev$ and characterized by both the substantial HNL mixing with the active neutrinos and the dipole portal coupling\cite{Gninenko:2009ks}, can avoid the corresponding bounds on $\mu_N$. This model appears, however, to be disfavored\cite{Brdar:2020tle} by the MINER$\nu$A data\cite{Park:2015eqa}. Such an explanation of the anomaly alone would also struggle to fully reproduce the angular distribution of the excess MiniBooNE events\cite{Jordan:2018qiy}. When presenting our results below, for completeness, we will mark the MinoBooNE RoI in the sensitivity reach plots in the dipole portal model, albeit we keep in mind the aforementioned constraints on the simplest such scenario.

\subsection{Dark gauge boson mediator\label{sec:darkgaugeboson}}

The secondary production of HNLs in neutrino scatterings can also be enhanced in the presence of light BSM mediators coupled to neutrinos. In particular, such a vector mediator $Z_D$ can naturally arise after gauging one of the global anomaly-free symmetries of the SM, although the relevant couplings are then suppressed, cf. Ref.\cite{Bauer:2018onh,Kling:2020iar}. In addition, as far as the upscattering cross section is concerned, $\nu (e/Z)\to N (e/Z)$, further suppression comes from the mixing parameter between the active neutrinos and HNLs, $U_{\nu N}^2$. This typically leads to negligible secondary production interaction rates in the allowed regions of the parameter space of such models. Instead, the relevant cross section can be much enhanced if the new dark gauge symmetry couples $Z_D$ directly only to the HNL, with a large allowed value of the dark coupling constant, $g_D\sim \mathcal{O}(1)$\cite{Ballett:2018ynz,Bertuzzo:2018itn}.

Additional couplings of $Z_D$ to quarks and electrons can appear e.g. due to a kinetic mixing or as a result of the mass mixing with the SM $Z$ boson. The latter possibility, however, is subdominant for $m_{Z_D}\lesssim\gev$. Instead, for such light $Z_D$, the kinetic mixing allows for larger interaction rates, with the upper limits dictated by the mixing parameter $\epsilon\lesssim 10^{-3}$, cf. Ref.\cite{Beacham:2019nyx} for a recent review of the relevant constraints.

The simplified model that incorporates both the kinetic mixing and the dark coupling of $Z_D$ to the HNL can be described by the following Lagrangian\cite{Ballett:2018ynz,Bertuzzo:2018itn}
\begin{equation}
\mathcal{L}_{D} \supset \frac{m_{Z_{D}}^{2}}{2} Z_{D \mu} Z_{D}^{\mu}+g_{D} Z_{D}^{\mu} \bar{N} \gamma_{\mu} N+e \epsilon Z_{D}^{\mu} J_{\mu}^{\mathrm{em}},
\label{eq:Ldarkgaugeboson}
\end{equation}
where the coupling of $Z_D$ to the electromagnetic current $J_{\mu}^{\mathrm{em}}$ is induced by the kinetic mixing. In this model, the upscattering cross section of the active neutrinos to HNLs scales with $g_D^2 U_{\nu N}^2\,\alpha\epsilon^2/m_{Z_D}^4$, and can be sizable for the dark gauge boson with the mass in the sub-GeV range. We provide the relevant expressions in \cref{app:sigma,app:decay}.

In order to reduce the number of free parameters of the model when presenting the results below, we vary the HNL mass and the dominant mixing parameter with the muon neutrino $\nu_\mu$, while we fix the other parameters. In particular, we focus on the benchmark scenario with fixed $\alpha_D=0.25$, $\alpha \epsilon^2=2\times 10^{-10}$, and $m_{Z_D}=30\MeV$, which was introduced in Ref.\cite{Bertuzzo:2018itn} to fit the MiniBooNE anomaly. In this case, for $m_N>m_{Z_D}$, the HNL produced in the active neutrino upscattering promptly decays into the on-shell mediator boson, $N\to\nu Z_D$. The latter subsequently decays inside FASER, $Z_D\to e^+e^-$, and generates a visible signature of two high-energy oppositely-charged tracks. The typical decay length of such $Z_D$ is given by
\begin{equation}
\bar{d}_{Z_D}\simeq (1~\textrm{m})\,\left(\frac{E_N}{300~\gev}\right)\,\left(\frac{30~\mev}{m_{Z_D}}\right)^2\,\left(\frac{2\times 10^{-10}}{\alpha \epsilon^2}\right).
\end{equation}

In addition, we also present the results for the case with the dominant mixing with the tau neutrinos and $m_N<m_{Z_D}$, which has recently been discussed in Ref.\cite{Jho:2020jfz} in the context of FASER searches. Here, the HNL undergoes a three-body decay, $N\to\nu e^+e^-$, via an intermediate off-shell $Z_D$. Such decays can then be detected based on the observation of $e^+$ and $e^-$ tracks in the FASER spectrometer. Importantly, in this case, the lifetime of the HNL can easily be large enough such that the dominant contribution to the signal rate in FASER comes from HNLs produced at the ATLAS IP, as discussed in Ref.\cite{Jho:2020jfz}.

In our study, we extended the previous analysis by checking the prospects of HNL searches with the electron scattering signature in FASER$\nu$2 due to both the inelastic $\nu_\tau e\to N e$ and the elastic $Ne\to Ne$ processes. We also take into account the additional production mode of the HNLs in decays of on-shell dark gauge bosons, $Z_D\to NN$.

Notably, the scattering cross section of the elastic scattering process is not suppressed by the mixing angle, $U_{\tau N}^2$. However, this suppression can occur in the production rate of $N$, which can be estimated from the tau neutrino production rate corrected for relevant helicity and phase-space factors, as dictated by the non-zero mass of the HNL\cite{Orloff:2002de}. As a result, both the elastic and inelastic contributions to the electron scattering signature can play a comparable role in the sensitivity reach plots. On the other hand, the suppression of $N$ production rate is no longer true for a very low mixing angle and small $m_N$. In this case, the HNL is dominantly produced in decays of the on-shell dark gauge bosons. As long as the HNL is sufficiently long-lived, this contribution to the signal rate of elastic scatterings off electrons is then independent of $U_{\tau N}^2$.

Last but not least, as we have already mentioned above, models employing light BSM vector or scalar mediators have also been proposed to explain the MiniBooNE anomaly. Both the scenarios with off-shell\cite{Ballett:2018ynz} and on-shell\cite{Bertuzzo:2018itn} mediators have been considered with a different HNL decay kinematics. The former model, however, is in strong tension\cite{Brdar:2020tle} with the T2K ND280 search for $2e$ tracks from HNL decays\cite{Abe:2019kgx}. This tension is somewhat less pronounced for the latter scenario, although it is also not favored by this and  the aforementioned MINER$\nu$A search\cite{Park:2015eqa}, by the CHARM-II data\cite{Vilain:1994qy}, and by the requirement to fit the angular distribution of the MiniBooNE events\cite{Arguelles:2018mtc}. For illustration purposes, in this study, we consider the model described by \cref{eq:Ldarkgaugeboson} with the minimum BSM particle content and we show the MiniBooNE RoI for this scenario following Ref.\cite{Bertuzzo:2018itn}. We note, however, that further discussion of this and other anomalies in more complex scenarios employing HNLs and either vector or scalar mediators have been proposed in the literature\cite{Abdullahi:2020nyr,Datta:2020auq,Dutta:2020scq,Abdallah:2020biq}. These could lead to even more rich FASER phenomenology. We leave these studies for future dedicated analyses.

\section{Results\label{sec:results}}

As discussed above, we have identified the simplified BSM models of our interest such that they illustrate a number of ways in which new physics signals could be observed in neutrino interactions in the FASER and FASER 2 multi-purpose detectors. Below, we discuss such opportunities in more detail and present the relevant sensitivity reach plots employing the detector designs introduced in \cref{sec:modeling}.

\subsection{Turning neutrinos into light with neutrino dipole portal to HNLs\label{sec:nutolight}}

As we have already mentioned in \cref{sec:dipoleportal}, the secondary production of HNLs in neutrino scatterings off electrons or nuclei via the dipole portal, can lead to a number of interesting phenomenological signatures in FASER. In particular, their subsequent decays, $N\to\nu\gamma$, inside the decay vessel or the ECC detector, can be observed as a clear excess of the single-photon-initiated EM showers after the BG rejection procedure is applied that we have outlined above.

\begin{figure}[t]
\begin{subfigure}{0.4\textwidth}
\centering
\includegraphics[scale=0.5]{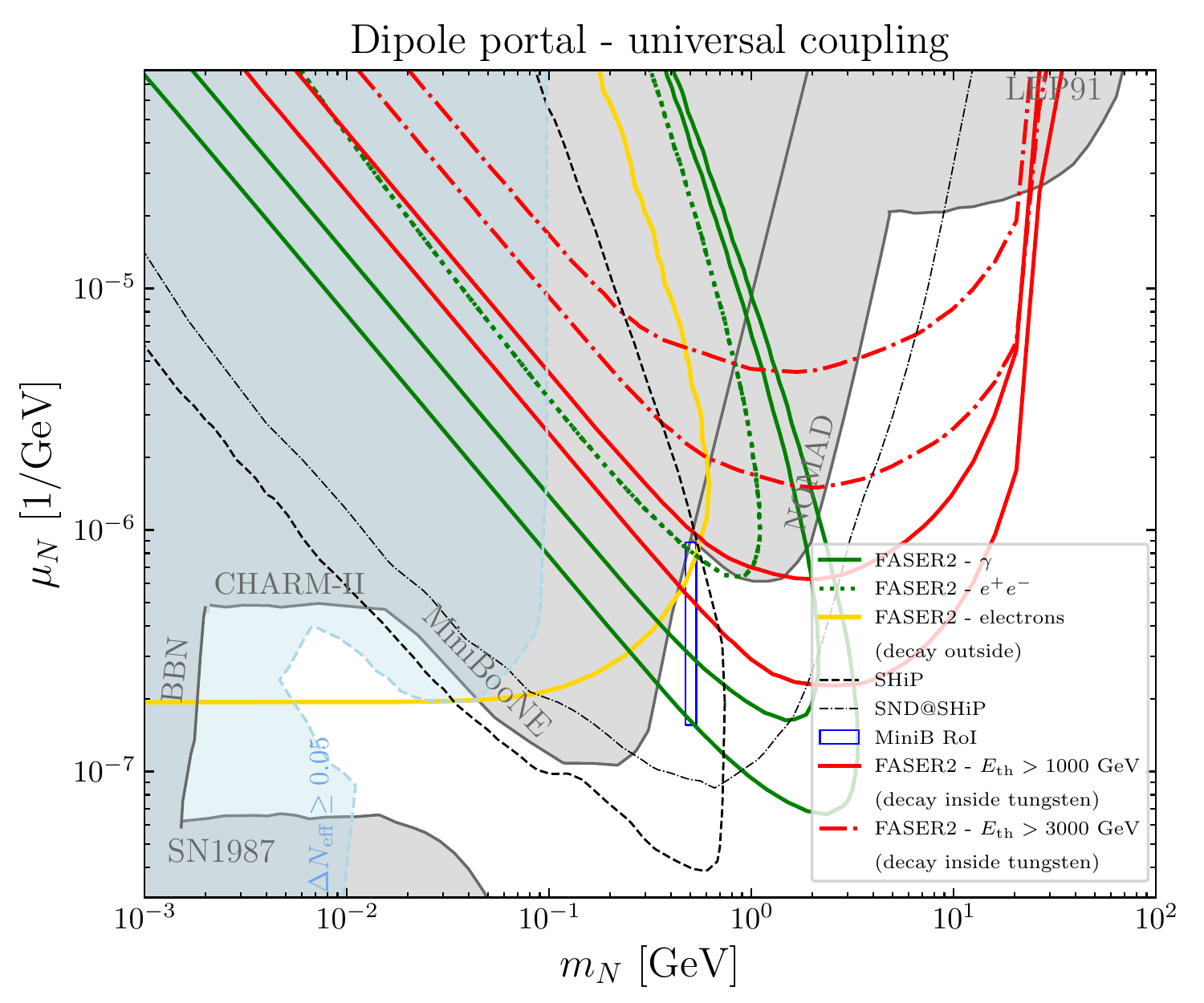}
\end{subfigure}
\hspace{1.2cm}
\begin{subfigure}{0.4\textwidth}
\centering
\includegraphics[scale=0.5]{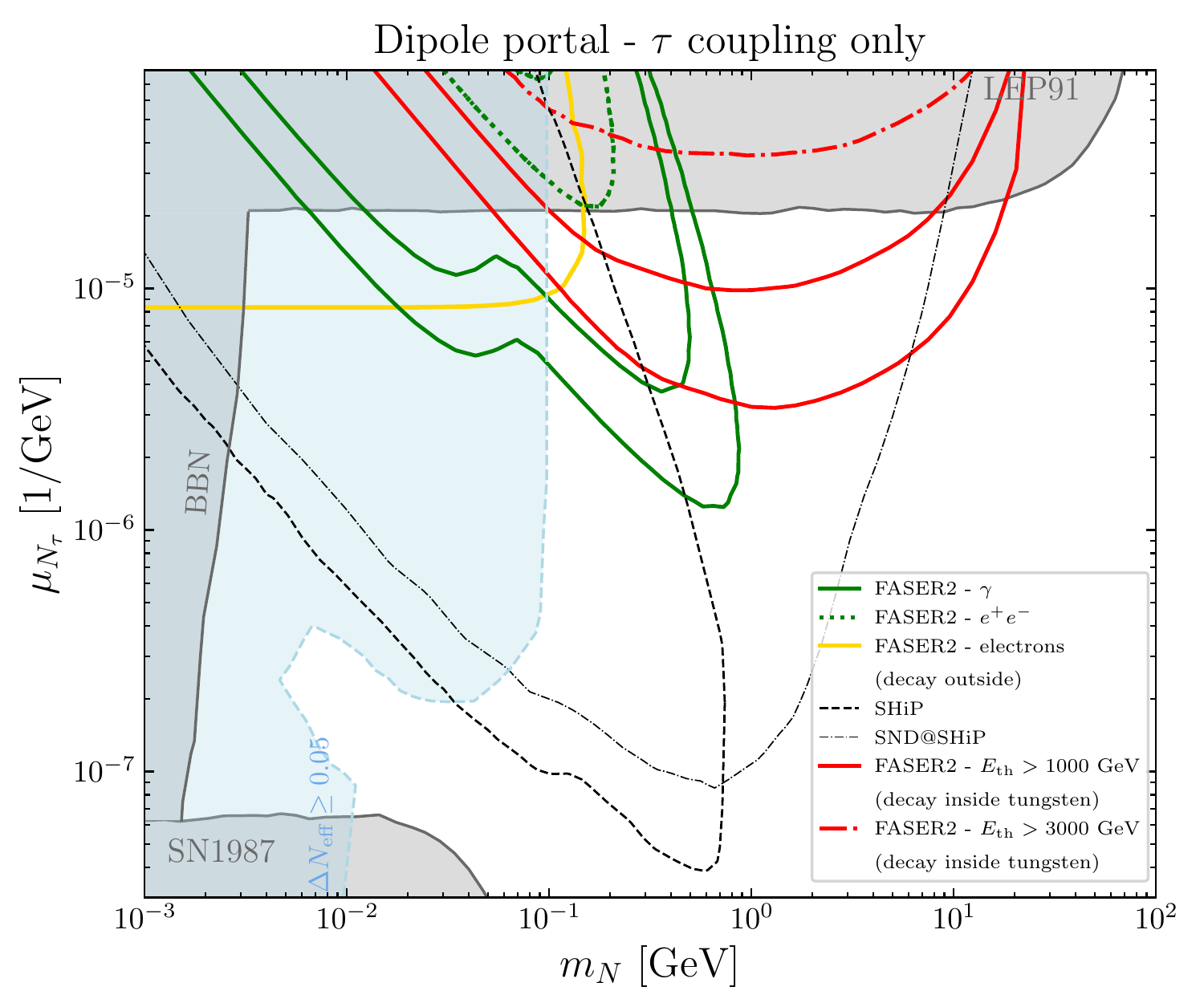}
\end{subfigure}
\caption{
The main result for the dipole portal model. In the left panel, we present the results for the model with the universal coupling to all of the neutrino flavors, while the right one corresponds to the model with the $\nu_\tau$-specific coupling. The green solid lines denote the expected number of the signal events in FASER$\nu$2 during HL-LHC that involve high-energy photons appearing inside the decay vessel ($N_{\textrm{ev}}=3,30$). The red solid (dash-dotted) lines correspond to such photons appearing in the ECC detector with the energy $E_\gamma>1~\tev$ ($E_\gamma>3~\tev$). The yellow solid line represents the expected exclusion bound from the search for the electron scattering events. The gray-shaded regions denote current bounds on both models (see the text for details). The dark blue rectangle denotes the regions of interest in the parameter space relevant for explaining\cite{Gninenko:2009ks,Magill:2018jla} the origin of the MiniBooNE anomaly\cite{AguilarArevalo:2007it,Aguilar-Arevalo:2020nvw}. The blue-shaded region corresponds to scenarios, in which the increased number of relativistic degrees of freedom in the early Universe predicted by the model could be used to relax the Hubble tension\cite{Brdar:2020quo}. In the left panel, we also show the expected reach of the proposed SHiP experiment focusing on the search for single photons in the decay vessel (black dashed line) or in the SND@SHiP detector (black dash-dotted line extended towards larger HNL masses) following Ref.\cite{Magill:2018jla}.
}
\label{fig:resultsdipole}
\end{figure}

\paragraph{Single photons in the decay vessel} In \cref{fig:resultsdipole}, we show with the green solid lines the results of such an analysis for the single photons appearing in the decay vessel of FASER 2. From the bottom to the top, the lines correspond to $N_{\textrm{ev}}=3$ and $30$ expected such events during the HL-LHC phase, while we have suppressed the lines corresponding to larger values of $N_{\textrm{ev}}$ for clarity of the plot. In the left panel, these results are presented for the dipole portal model with a universal coupling $\mu_N$ to all the neutrino flavors, while the right panel corresponds to the scenario with the $\nu_\tau$-specific coupling $\mu_{N\tau}$. In the former case, for comparison, we also show the expected reach of the SHiP experiment following Ref.\cite{Magill:2018jla}.

In the plots, the gray-shaded region corresponds to the current experimental bounds. The dominant such bounds come from null searches in the CHARM-II\cite{Geiregat:1989sz}, MiniBooNE\cite{AguilarArevalo:2007it} and NOMAD\cite{Altegoer:1997gv} experiments, as well as from the LEP search for the $\gamma+\slashed{E}_T$ signature\cite{Abreu:1996vd}. For light and very weakly coupled HNLs, the cosmological and astrophysical constraints from the Big Bang Nucleosynthesis (BBN) and observations of Supernova SN1987A become important. We follow Refs\cite{Coloma:2017ppo,Magill:2018jla,Shoemaker:2018vii,Brdar:2020quo} when presenting these bounds. Notably, in the $\nu_\tau$-specific scenario, with negligible couplings to electron and muon neutrinos, the constraints from the past $\nu_\mu$ experiments cease to be valid. On the other hand, for the center-of-mass energy available at the LHC, tau neutrinos can be efficiently produced via the decay of the charm mesons and tau leptons. This allows one to probe this scenario in FASER 2.

We find that up to $\mathcal{O}(10^3)$ HNL-induced single-photons can be observed for the model with the universal coupling in the allowed region of the parameter space and about $100$ such events are expected in the $\nu_\tau$-specific scenario. Instead, both the lower luminosity and the size of the detector, results in no more than $\mathcal{O}(10)$ of single-photon events expected in FASER during LHC Run 3. This could allow to start probing only a small region of the parameter space of the model with the universal coupling $\mu_N$ in the initial run of the experiment.

Importantly, as mentioned in \cref{sec:dipoleportal}, the energy spectrum of photons from HNL decays is dictated by the interplay between the relevant decay length and the distance the HNLs need to travel between the tungsten plates in the neutrino detector and the decay vessel. In particular, for short-lived HNLs, this spectrum is shifted towards higher energies resulting in larger boost factors, which makes it easier to disentangle the signal events from the muon-induced BG. The latter is expected to peak at smaller energies due to growing bremsstrahlung cross section of soft photons\cite{Groom:2001kq}.

\begin{figure}[t]
\centering
\includegraphics[scale=0.49]{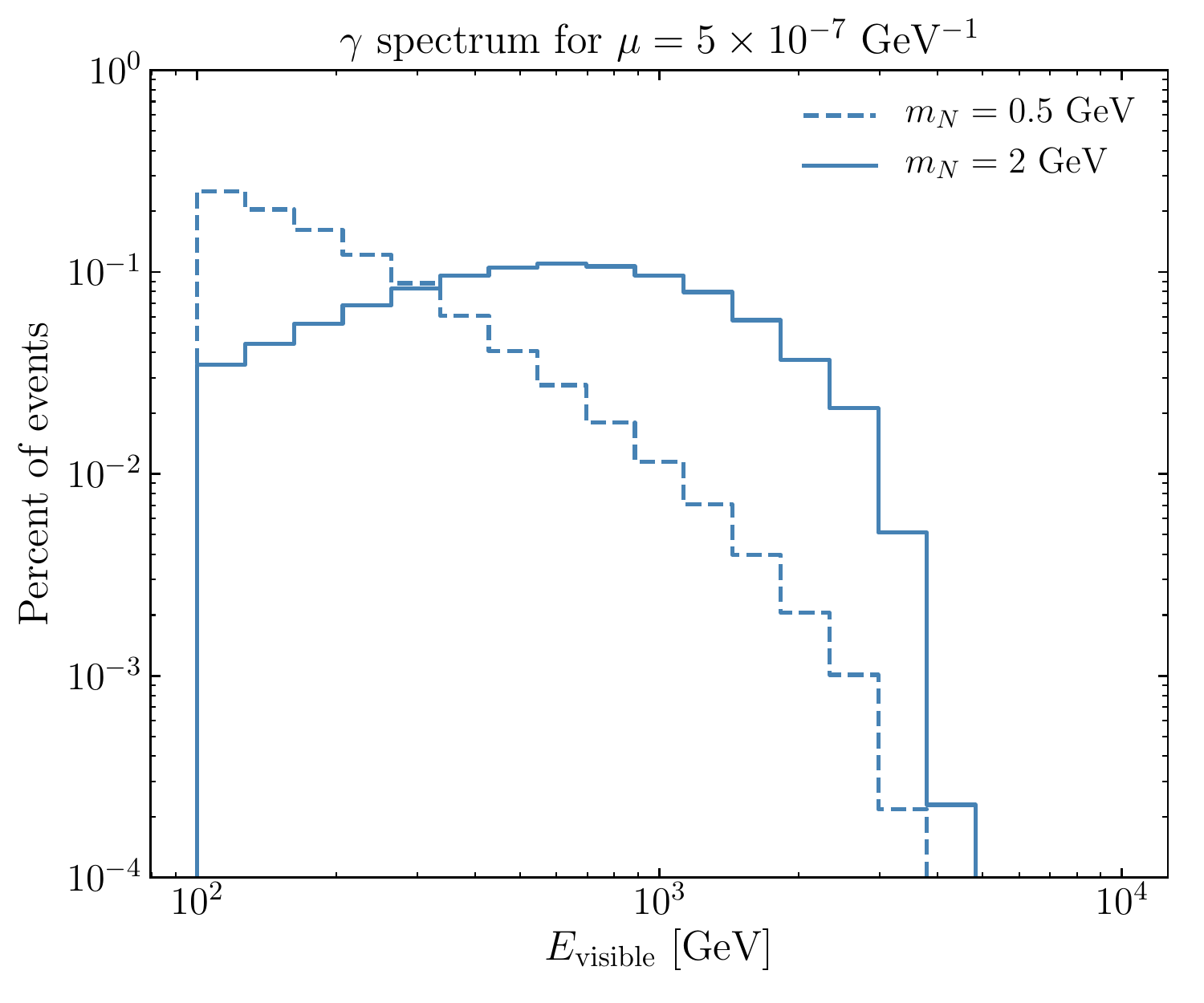}
\hfill
\includegraphics[scale=0.49]{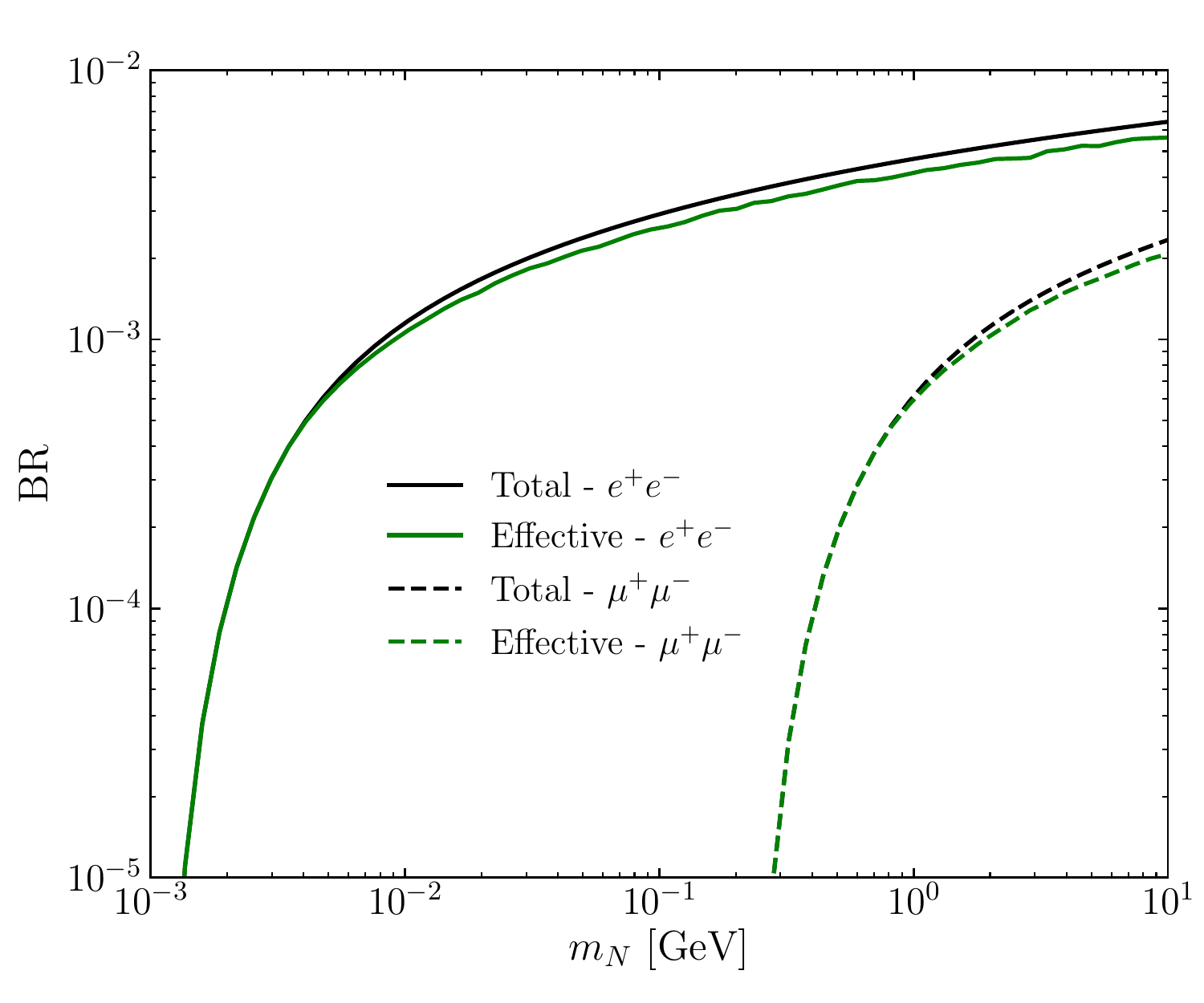}
\caption{The results for the model with the dipole portal to HNLs and the universal coupling $\mu_N$. \textsl{Left}: The energy spectrum of photons coming from the HNL decays in the FASER decay vessel for the benchmark values of the model parameters as indicated in the plot. \textsl{Right}: The branching fraction into three-body decays, $N\to\nu(\gamma^\ast\to\ell\ell)$, where $\ell=e$ (solid black line) or $\mu$ (dashed black), as a function of the HNL mass $m_N$. The green solid (dashed) lines correspond to a slightly lower ``effective'' branching fraction that takes into account the analysis cut on the visible energy, $E_{\textrm{vis}}>100~\gev$, for the electrons (muons) in the final state. These lines assume that the parent HNL energy is equal to $E_N=1~\tev$.}
\label{fig:gammaspectrum}
\end{figure}

We show such a photon energy spectrum for the BSM signal events in the left panel of \cref{fig:gammaspectrum} for the two benchmark scenarios within the reach of FASER 2. They correspond to the two different HNL masses, $m_N=0.5$ or $2~\gev$, but to the same value of the coupling constant equal to $\mu_N=5\times 10^{-7}$. As can be seen, for the larger value of the HNL mass, the BSM signal, indeed, peaks around $E_\gamma\sim \tev$. Instead, for $m_N=0.5~\gev$, new-physics-induced photons can be less energetic, as dictated by the smaller value of the HNL lifetime. They would have to be searched for as an excess over the residue softer muon-induced BG, after the rejection procedure is applied. The procedure is based on the parent muon detection and the lack of the photon showering in the upstream parts of the detector. We note, however, that such BG should still be suppressed for the case with $E_\gamma>100~\gev$ that we focus on. Instead, we leave for future studies and detector simulations, a detailed analysis of the remaining irreducible BG from the high-energy electron neutrino interactions in the preshower detector or in the upstream part of the calorimeter with the possible backsplash effect in place.

In \cref{fig:resultsdipole}, with the dotted green lines, we also show the expected sensitivity reach of FASER 2 in the search for an $e^+e^-$ pair from the three-body decay $N\to\nu(\gamma^\ast\to\ell\ell)$. The relevant branching fraction is suppressed by $2$ to $3$ orders of magnitude with respect to the dominant two-body decay, as shown in the right panel of \cref{fig:gammaspectrum}. In the plot, we present both the total branching fraction, and the ``effective'' one, which takes into account the lower threshold for the visible energy in the signal events assuming that the parent HNL energy is equal to $E_N=1~\tev$. Unfortunately, the sensitivity reach of FASER 2 in this search will cover only a very small region of the parameter space of the model, so it would not be complementary to the search for single photons.

\paragraph{Single photons in the neutrino ECC detector} After the secondary production process, the HNL decays into single photons can also happen inside the ECC detector FASER$\nu$2. The search for such high-energy photons with $E_\gamma>1~\tev$ could play a complementary role to the similar signature discussed above based on the photons appearing in the decay vessel. This complementarity primarily concerns HNLs with even smaller lifetimes that decay promptly after the production in coherent scatterings off nuclei, cf. \cref{sec:modeling} for the relevant discussion about the expected BG in this search. We note that a similar search has already been discussed in the context of the proposed SND@SHiP detector in Ref.\cite{Magill:2018jla}.

We show the contours with a fixed number of such events, $N_{\textrm{ev}}=3$ and $30$, for the $1~\tev$ ($3~\tev$) photon energy threshold with the red solid (dash-dotted) lines in \cref{fig:resultsdipole} for the FASER$\nu$2 experiment during HL-LHC. As can be seen, in both scenarios of our interest, this search would extend the relevant reach of FASER$\nu$2 and SHiP towards the larger values of both the HNL masses and the coupling constants. Assuming the $1~\tev$ threshold, up to about $10^4$ and $100$ such events in FASER$\nu$2 during HL-LHC are expected for the scenario with the universal coupling $\mu_N$ and the $\nu_\tau$-specific coupling $\mu_{N\tau}$, respectively. Notably, in the former model, one could expect a few hundred signal events even for the $3~\tev$ photon energy threshold. In some cases, the detectable event rates are obtained for both the aforementioned search for single photons in the decay vessel and the similar signature inside the ECC detector. This increases the combined discovery prospects of FASER 2 in such scenarios.

\subsection{The bi-modal $e^+e^-$ spectrum in the search for HNLs and dark gauge bosons\label{sec:bimodal}}

While the $e^+e^-$ signature is suppressed in the dipole portal case, it remains the main discovery channel of HNLs in the model with the dark gauge boson mediator $Z_D$, cf. \cref{sec:darkgaugeboson}. In particular, in the case of $m_{Z_D}<m_{N}$, which we first focus on, the HNLs decay promptly and primarily into the invisible final state, $N\to\nu Z_D$. The subsequent decays of the dark gauge boson, $Z_D\to e^+e^-$, in the FASER decay vessel, can lead to visible signatures with $E_{e^+e^-}>100~\gev$.

\begin{figure}[t]
\includegraphics[scale=0.45]{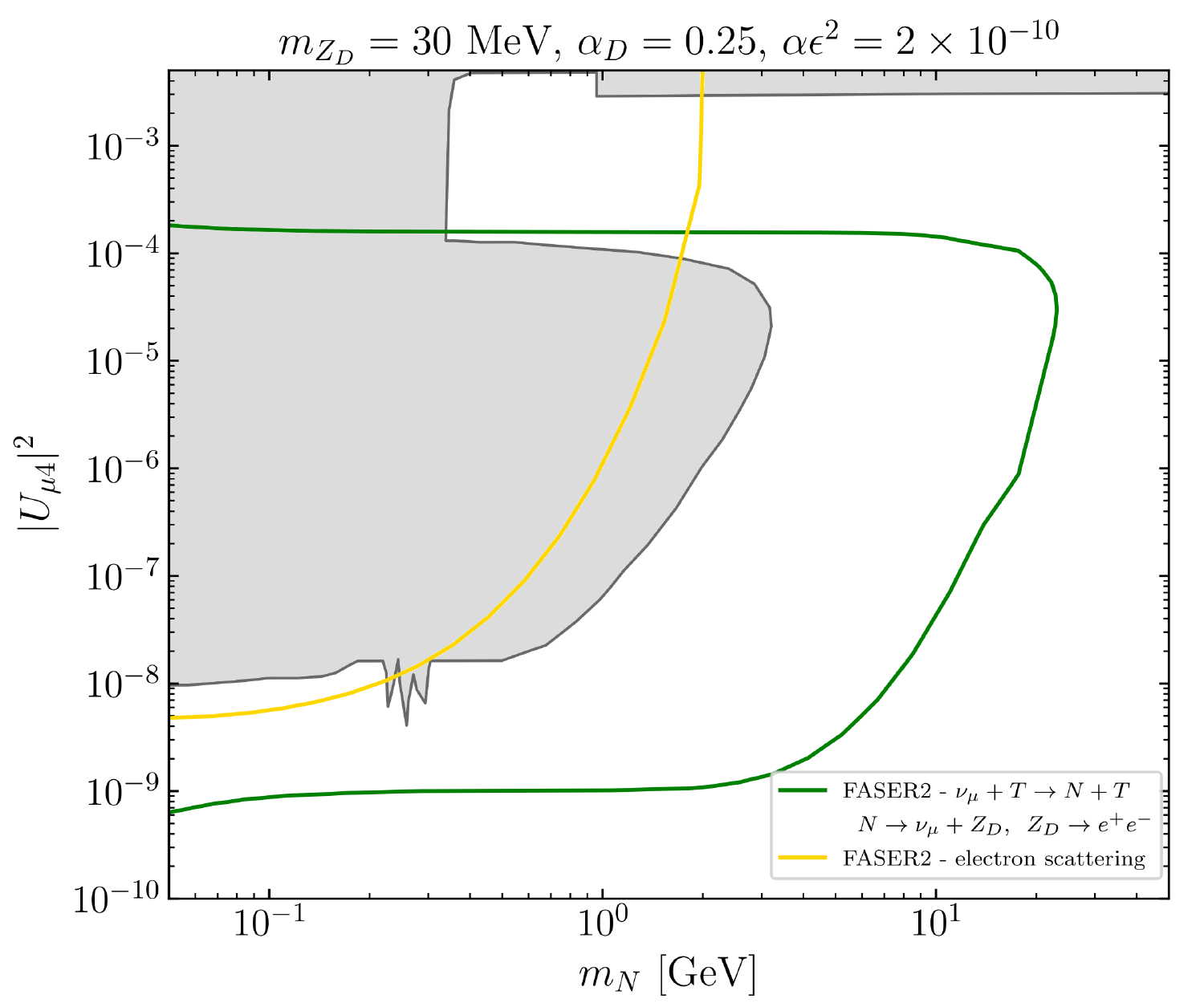}
\hfill
\includegraphics[scale=0.45]{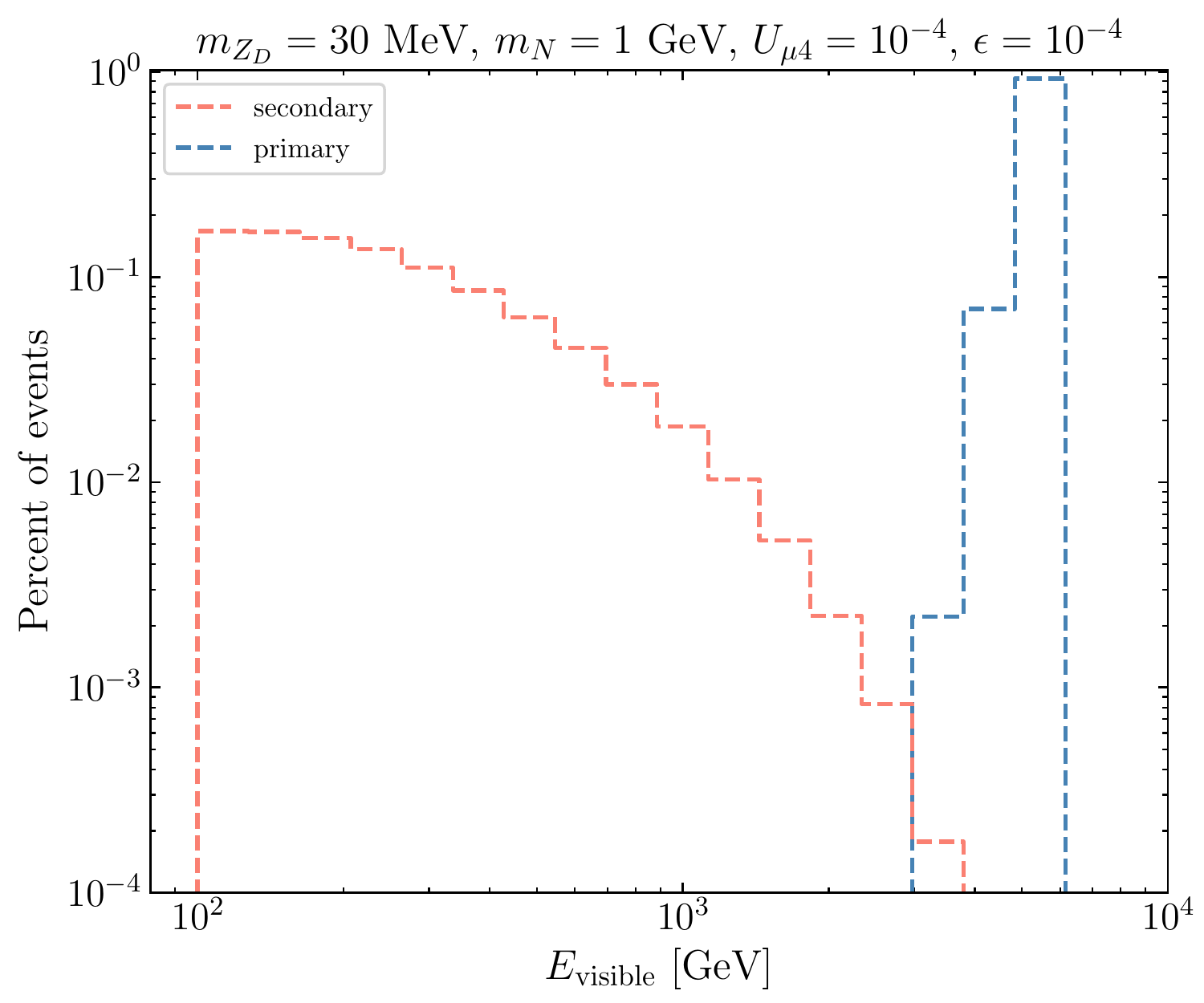}
\caption{\textsl{Left}: The sensitivity reach of FASER 2 in the model with the dark gauge boson mediator between the HNLs and SM particles. We vary the mass and the mixing angle of the HNL, while other parameters of the model are fixed, as shown in the plot. The gray-shaded region corresponds to the currently excluded region in the parameter space of the model (see the text for details). The FASER 2 sensitivity reach based on the $e^+e^-$ pair detection from the sequential process, $\nu Z\to Z (N \to \nu Z_D)$ with the subsequent decay $Z_D\to e^+e^-$, is shown with the green solid line. The additional sensitivity reach in this model is associated with a direct $Z_D$ production at the ATLAS IP. This is not shown in the plot, as it does not depend on the mixing angle and mass of the HNL. The yellow solid line corresponds to the expected sensitivity reach of FASER$\nu$2 based on the electron scattering signature. \textsl{Right}: The energy spectrum of $e^+e^-$ pairs from $Z_D$ decays. The blue histogram corresponds to the primary production of $Z_D$s at the ATLAS IP, while the orange one is relevant for the secondary production of the dark gauge boson as a result of the neutrino upscattering in front of the detector. The parameters of the model chosen to prepare both histograms are indicated in the plot.
}
\label{fig:resultsdarkgaugeboson}
\end{figure}

In \cref{fig:resultsdarkgaugeboson}, with the solid green line, we present the relevant expected sensitivity reach of FASER 2 in the $(m_N,U_{\mu N})$ plane, where $U_{\mu N}$ is the HNL mixing angle with the active muon neutrino. In the plot, we fix the values of the other parameters of the model: $m_{Z_D}=30~\mev$, $\alpha_D=0.25$ and $\alpha\epsilon^2 = 2\times 10^{-10}$. This choice of the parameters could be further motivated by the proposal to fit the MiniBooNE anomaly in this scenario\cite{Bertuzzo:2018itn}, cf. the discussion in \cref{sec:darkgaugeboson}. The dominant bounds on the model for $m_N\lesssim \mathcal{O}(1~\gev)$ come from searches by the MINER$\nu$A\cite{Park:2015eqa} and CHARM-II\cite{Vilain:1994qy} experiments, and we implement them following Ref.\cite{Arguelles:2018mtc}.\footnote{We note that, while the CHARM-II bounds for $m_N\leq 1~\gev$ have been obtained with the dedicated MC simulation in Ref.\cite{Arguelles:2018mtc}, for heavier HNLs we extrapolate these bounds using only simple recasting procedure.} Additional relevant constraints are associated with rare meson decays, the muon decay Michel spectrum, and lepton universality\cite{Atre:2009rg,deGouvea:2015euy}, as discussed in Ref.\cite{Bertuzzo:2018itn}. A recent review and update of the constraints on the HNLs can also be found in Ref.\cite{Bolton:2019pcu}. Notably, the lack of the FASER sensitivity and the CHARM-II and MINER$\nu$A bounds in the part of the parameter space corresponding to large values of mixing angle, $U_{\mu N}^2\sim 10^{-3}$, is due to the dominant invisible decays of the dark gauge boson in this case, $Z_D\to\nu\bar{\nu}$.

In the plot, we show the reach of FASER 2 corresponding to only the secondary production of HNLs and $Z_D$s in front of and inside the detector. As can be seen, FASER 2 could probe currently unexplored regions in the parameter space corresponding to low values of the mixing angle and to the increasing HNL mass, up to $m_N\sim \mathcal{O}(10~\gev)$. In the case of light HNLs, FASER 2 will also cover the MiniBooNE RoI. For such light HNLs, FASER 2 will also provide complementary probe of this scenario to the search for HNLs produced in rare kaon decays in the NA62 experiment, cf. Ref.\cite{Ballett:2019pyw}.

Importantly, although relatively short-lived $Z_D$s that we study struggle to survive the entire distance between the ATLAS IP and FASER without decaying, there remains a small fraction of them produced in $pp$ collisions that can contribute to the signal rate in the detector. This is complementary to the dark gauge bosons produced in neutrino scatterings. The primary production of $Z_D$ at the ATLAS IP, however, does not depend on the mixing angle $U_{\mu N}$ and mass of the HNL. It then corresponds to a fixed expected number of signal events, which adds to the events indicated in the reach plot in the left panel of \cref{fig:resultsdarkgaugeboson}.

The interplay between the two production processes is driven on the one side by the exponentially suppressed decay-in-volume probability for short-lived $Z_D$s produced at the ATLAS IP. On the other hand, it depends on a small value of the secondary production cross section in front of the decay vessel. The former suppression results in the spectrum of $Z_D$ from the primary production that are peaked towards larger energies, $E_{Z_D}\gtrsim\tev$, so that they can reach FASER without decaying. Instead, the dark gauge bosons produced in the secondary production processes in the vicinity of the detector, favor lower energies of the visible signal in the detector. In this case, while the energy spectrum of parent neutrinos is peaked around a few hundred $\gev$, on average only about half of this energy goes into $Z_D$ after the HNL decay. This effectively suppresses a typical energy of the dark gauge boson to $E_{Z_D}\sim 100-300~\gev$. We illustrate this interesting phenomenological effect in the right panel of \cref{fig:resultsdarkgaugeboson}, in which a bi-modal energy spectrum of $e^+e^-$ pairs detected in FASER 2 is shown, with each of the peaks corresponding to the same BSM species but a different production process of $Z_D$.

\subsection{Probing light HNLs via scatterings off electrons\label{sec:scatoffelectrons}}

The BSM interactions of neutrinos can also manifest themselves in the FASER$\nu$2 detector in enhanced scatterings off electrons, cf. Ref.\cite{Batell:2021blf} and \cref{sec:modeling} for a more detailed discussion about the expected BG in this case. We present the relevant expected future exclusion bounds for the dipole and dark gauge boson portals in \cref{fig:resultsdipole,fig:resultsdarkgaugeboson} with the solid yellow lines.

The results shown there correspond to the neutrino upscattering events, $\nu e\to N e$. Given the typical incident neutrino energy of order several hundred $\gev$, the center-of-mass energy of such collisions allows one to produce HNLs with the mass $m_N\lesssim\gev$. In particular, in the case of the dipole portal model, the complementarity between the scattering off electrons and both of the aforementioned single-photon signatures, allows one to probe a broad range of the HNL masses, $1~\mev\lesssim m_N\lesssim 10~\gev$. Interestingly, in this model, the scattering signature itself can probe parts of the parameter space that correspond to increased values of the effective number of relativistic degrees of freedom in the early Universe, $0.05 \lesssim \Delta N_{\textrm{eff}}\lesssim 0.3$, as also shown in \cref{fig:resultsdipole}. On the one hand, this region would be in tension with a conservative bound derived on $\Delta N_{\textrm{eff}}$ based on the cosmic microwave background observations and studies of baryon acoustic oscillations\cite{Philcox:2020vvt}. On the other hand, though, such increased $\Delta N_{\textrm{eff}}$ could relax the so-called Hubble tension\cite{Bernal:2016gxb,Knox:2019rjx}. Hence, we present the relevant region in the parameter space with a blue shaded region in \cref{fig:resultsdipole} following the discussion in Ref.\cite{Brdar:2020quo}.

\begin{figure}[t]
\includegraphics[scale=0.49]{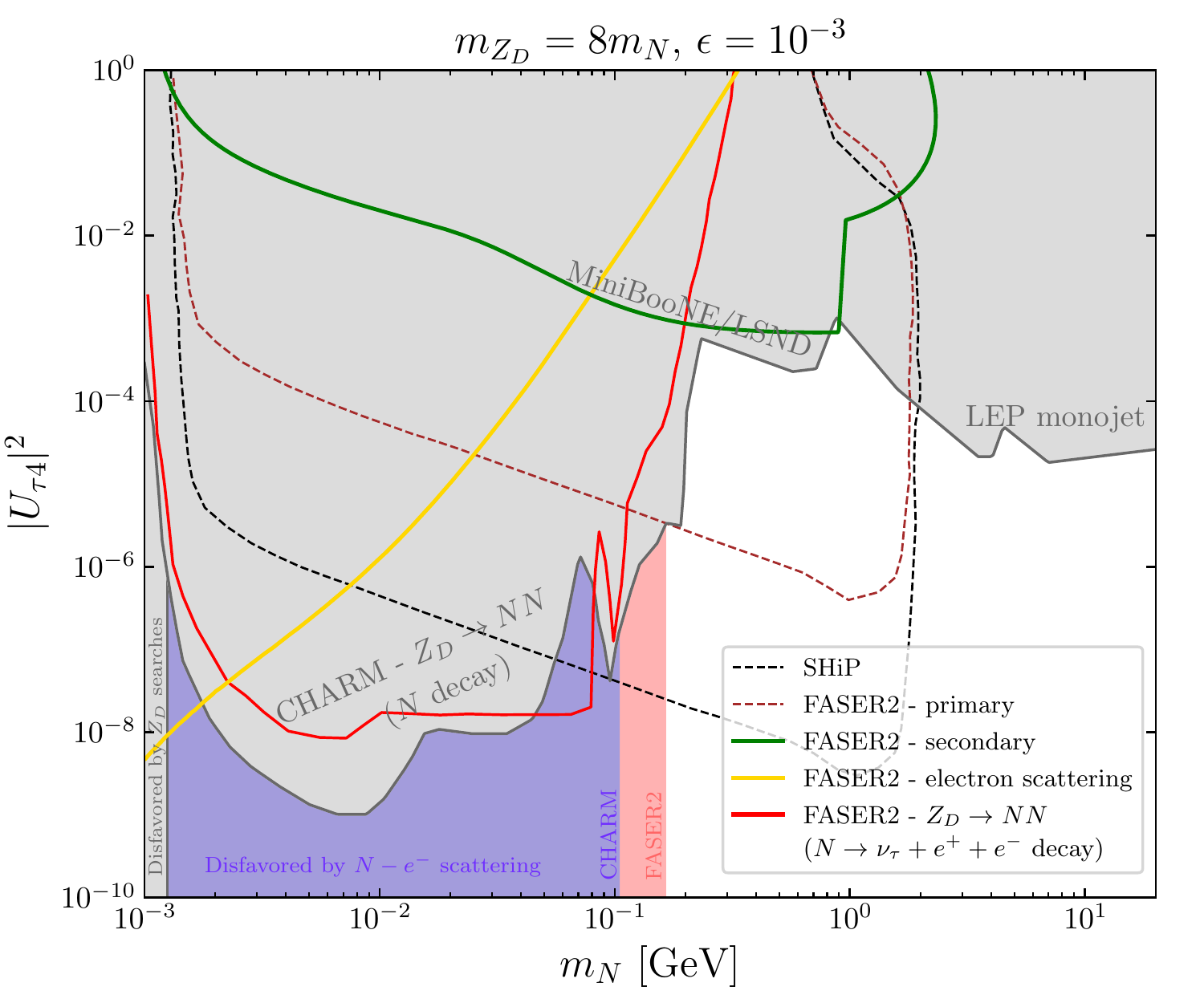}
\hfill
\includegraphics[scale=0.49]{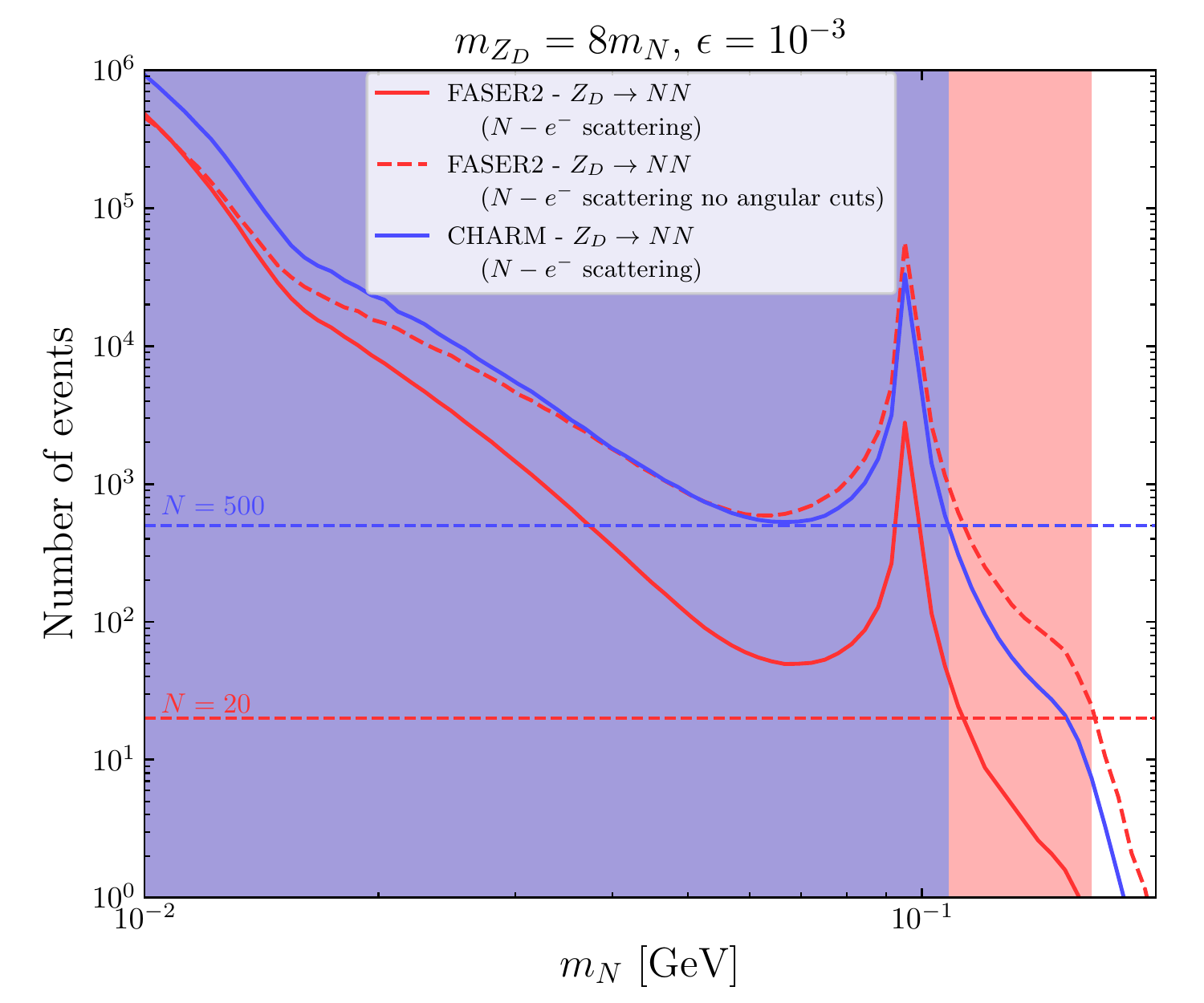}
\caption{The results for the model with the dark gauge boson heavier than the HNL, $m_{Z_D}=8 m_N$, and with the dominant mixing between the HNLs and active tau neutrinos, $U_{\tau N}$. The kinetic mixing parameter in the plots is also fixed to be equal to $\epsilon=10^{-3}$. \textsl{Left}: The FASER 2 and SHiP sensitivity in the $(m_N,U_{\tau N})$ plane. The dashed red and blue lines for both the experiments, respectively, are shown following Ref.\cite{Jho:2020jfz}. The additional FASER 2 sensitivity due to the secondary production of the HNL is shown with the green solid line. The red solid line corresponds to the additional FASER 2 reach due to the possible production of the HNLs in the dark gauge boson decays, $Z_D\to NN$. The yellow solid line is the expected exclusion bound relevant for the search for inelastic, $\nu e\to Ne$, and elastic, $Ne\to Ne$, scatterings off electrons, assuming that in the latter case the HNLs are produced due to the mixing with the active neutrinos. The relevant results for FASER 2 for the HNLs produced in the $Z_D$ decays are shown with the red-shaded region, while the past bounds from the CHARM-II experiments are indicated with the blue-shaded region. \textsl{Right}: The expected number of elastic scattering events, $Ne\to Ne$, in the CHARM-II (blue solid line) and FASER 2 (red solid line) that take into account the relevant cuts for the electron recoil energy and angle. The red dashed line corresponds to the expected events in FASER 2 after the recoil angle cuts have been relaxed. The dashed horizontal lines correspond to $N_{\textrm{ev}}=20$ (red) or $500$ (blue) events, and are used to set the CHARM-II and FASER 2 bounds (see the text for details).
}
\label{fig:darkgaugeboson2}
\end{figure}

Other interesting phenomenological aspects of the scattering signature appear in the model with the dark gauge boson mediator heavier than the HNL, $m_N<m_{Z_D}$. This scenario has also recently been discussed in Ref.\cite{Jho:2020jfz}. We present the relevant result in \cref{fig:darkgaugeboson2} assuming $m_{Z_D} = 8 m_N$ and the dominant coupling of $Z_D$ to the tau neutrinos. Here, we also assume the fixed values of the couplings constants $g_D=1$ and $\epsilon=10^{-3}$. Notably, this value of the kinetic mixing parameter lies close to the upper bound from the BaBaR search for invisible decays of the dark photon. In this case, the BaBaR search constrains $Z_D\to N N$ decays into the long-lived HNLs that often leave the detector before further decaying. The other dominant bounds, which are shown in the plot, come from the searches for $\tau$ lepton decays in BaBaR and Belle-II\cite{Kobach:2014hea}, the LEP monojet searches\cite{Abreu:1996pa}, and the past beam-dump experiments CHARM-II\cite{Orloff:2002de} and NOMAD\cite{Astier:2001ck}. We implement them following Ref.\cite{Jho:2020jfz}, beside the CHARM-II bound, which we update by taking into account the additional flux of HNLs from the $Z_D\to N N$ decays and the constraint from the elastic scattering processes, $Ne\to Ne$, in the detector, as discussed below.

We note that while the aforementioned benchmark scenario has been chosen only for illustrative purposes, similar number of scattering events is expect in the detector for other values of the model parameters as long as the combination $g_D^2 U_{\nu N}^2 \alpha \epsilon^2/m_{Z_D}^4$ is kept constant and one assumes $m_{Z_D}>2 m_N$ such that $Z_D\to NN$ decays are kinematically allowed. For $m_{Z_D}\sim m_N$, we also expect additional bounds from direct $Z_D$ searches for low dark vector masses, $m_{Z_D}\lesssim 10~\mev$. This is only mildly present in our reach plots, since we assume $m_{Z_D} = 8 m_N$. On the other hand, further increasing $m_{Z_D}$ would generally only suppress the expected number of events since there is no room for a similar increase in $\epsilon$ to compensate for larger $m_{Z_D}$. This is, again, due to the constraints from direct searches for $Z_D$.

The sensitivity reach of FASER$\nu$2 in the scattering signature in this model receives contributions from the inelastic processes, $\nu e\to N e$, as well as from the elastic processes, $N e\to N e$.\footnote{Instead, the inelastic processes $N e\to \nu e$ play a subdominant role with respect to the other two contributions, as it is suppressed with respect to them by the additional square of the mixing angle in either the HNL production or scattering processes.} As we have already mentioned above, the latter scattering cross section does not depend on the mixing angle between the HNL and active tau neutrino, $U_{\nu\tau}$, and the only such dependence can be in the HNL production. It can then provide a similar contribution to the signal rate to the inelastic scattering of the active neutrinos. We show the sum of both such contributions with the yellow line in the left panel in \cref{fig:darkgaugeboson2}.

On the other hand, since the HNLs are also produced in the decays of on-shell dark gauge bosons, this adds to the total scattering signal rate in a way that is fully independent of the mixing angle.\footnote{One should only require that $U_{\tau N}$ is small enough, such that the HNL does not decay before reaching the detector.} We, therefore, treat this contribution separately in the reach plot in the $(m_N,U_{\tau N})$ plane in the left panel of \cref{fig:darkgaugeboson2}. In the right panel of \cref{fig:darkgaugeboson2}, we show the expected number of such signal events as a function of the HNL mass for both the CHARM-II and FASER$\nu$2 experiments. Given the total observed number of $\nu$ and $\bar{\nu}$ electron scattering events in CHARM, which is equal to $2677 + 2752$\cite{Geiregat:1991md}, and using the relevant flux uncertainties of order $5\%$\cite{Vilain:1992wx}, we expect that a few hundred of additional scattering events in CHARM from $Ne\to Ne$ processes would be a clear indication of the BSM effect. For the illustration purposes, we then assume that scenarios predicting $N_{\textrm{ev}}\gtrsim 500$ such events are already excluded. This analysis takes into account the cuts on the electron recoil energy and angle used by the CHARM collaboration, which is implemented following Ref.\cite{Arguelles:2018mtc}.

The similar such cuts designed to search for new physics effects in electron scattering events in FASER$\nu$2\cite{Batell:2021blf} lead to much suppressed number of neutrino-induced BG events, $N_{\textrm{ev}}\sim 10$ or $\mathcal{O}(100)$, where the latter estimate neglects the cuts on the electron recoil angle. This leaves a room for improved constraints on this scenario based on the observation of the $Ne\to Ne$ events in FASER$\nu$2. It is especially the case for the increasing mass of $Z_D$, which is dominantly produced in the bremsstrahlung process, $pp\to ppZ_D$, cf. \cref{sec:fluxandprod} and Ref.\cite{Feng:2017uoz} for the discussion relevant for the far-forward region of the LHC. In the right panel of \cref{fig:darkgaugeboson2}, we show the relevant number of events as a function of $m_{N} = m_{Z_D}/8$ for a fixed value of the kinetic mixing parameter $\epsilon=10^{-3}$. While the resulting HNL spectrum is hard with the typical energy $E_N>100~\gev$, the spectrum of recoiled electrons after the scattering is shifted towards lower energies, cf. Ref.\cite{Batell:2021blf} for an extensive similar discussion about dark matter scattering off electrons in the model with light vector mediators. For illustration purposes, in the plots, we set the expected future exclusion bounds based on the observation of more than $20$ neutrino-like BSM elastic scattering events, $Ne\to Ne$. This is indicated with the red-shaded region in \cref{fig:darkgaugeboson2}. As can be seen, this search extends the expected FASER$\nu$2 sensitivity towards larger HNL masses with respect to the CHARM-II bounds indicated with the blue-shaded region.

\section{Conclusions\label{sec:conclusions}}

The neutrinos remain among the least experimentally tested SM species. This is especially the case at high energies, where the properties of many other known particles have already been thoroughly investigated in colliders. The high-energy neutrino physics, instead, provides a potentially very rewarding field of BSM research, which awaits a detailed exploration in future dedicated searches.

The recently approved FASER experiment, and its possible successor FASER 2, along with their neutrino subdetectors FASER$\nu$ and FASER$\nu$2, will pave the way for such an exploration beginning from LHC Run 3 throughout the HL-LHC phase. The experiments were originally proposed to search for light and long-lived BSM particles\cite{Feng:2017uoz,Ariga:2018zuc}, as well as to study the interactions of neutrinos at $\tev$ energies\cite{Abreu:2019yak}. Since then, however, several other physics motivations have been introduced for BSM searches along the beam collision axis in the far-forward region of the LHC. These include i.a. the search for DM scattering events\cite{Batell:2021blf} or for signatures of milli-charged particles\cite{Foroughi-Abari:2020qar}. In this study, we analyze the corresponding discovery potential for new physics particles appearing in high-energy neutrino interactions. While we have discussed the capabilities of both the FASER and FASER 2 experiments, a particular emphasis has been put on the larger successor detector to take data during the HL-LHC era.

We have shown that such searches, besides employing a specific location of the experiment, could much benefit from the unique properties of the FASER 2 detector. This also includes a possible interplay between its spectrometer and the neutrino subdetector. In particular, the latter allows one for a very precise reconstruction of interaction vertices. This leads to several possible signatures that can be used to study new physics. We have discussed the discovery prospects based on the standard search for two high-energy oppositely-charged tracks, but also the search for high-energy photons appearing in the detector and the single-electron scattering signature. They provide complimentary discovery channels and significantly extend the sensitivity reach in certain BSM scenarios.

The new physics models analyzed in our study have been chosen for the best illustration of this rich phenomenology. To this end, we have focused on BSM scenarios with $\gev$-scale heavy neutral leptons that can be produced in the active neutrino scatterings inside or in front of the detector. This allows FASER to effectively work as the high-energy neutrino beam-dump experiment. We also point out the interesting effects that can appear due to a combination of different production modes of new physics species.

While we focus on the simplest BSM scenarios employing the SM photon or dark gauge boson mediators between the neutrinos and other SM species, we have also stressed that other similar, though possibly more complex, models have been considered in the literature. This has been done especially in connection to the persisting MiniBooNE anomaly\cite{AguilarArevalo:2007it,Aguilar-Arevalo:2020nvw}, cf. Refs\cite{Abdullahi:2020nyr,Datta:2020auq,Dutta:2020scq,Abdallah:2020biq}.

Last but not least, besides direct production of new physics species, other signatures of BSM physics can also manifest itself in high-energy neutrino interactions in the FASER and FASER 2 detectors. This could be e.g. due to oscillations to sterile neutrinos\cite{Abreu:2019yak}, the double-bang signature in the emulsion detector\cite{Jodlowski:2019ycu}, or the enhanced neutrino trident production in the presence of light mediators, cf. a general discussion in Ref.\cite{Altmannshofer:2014pba}. We leave the further detailed analysis of these effects for the future dedicated studies for FASER. We also note that a similar research agenda would be relevant for the recently proposed Forward Liquid Argon Experiment (FLArE)\cite{Batell:2021blf} to be placed in front of FASER 2 in the Forward Physics Facility (FPF)\cite{SnowmassFPF} during the HL-LHC era. The advent of precision high-energy neutrino physics at the LHC opens up a new window to improve our understanding of these elusive particles and their possible connections to a more fundamental BSM description of microscopic interactions in nature.

\medskip

\paragraph{Acknowledgements} We would like to thank Akitaka Ariga, Tomoko Ariga and Iftah Galon for useful discussions. We thank Felix Kling and Leszek Roszkowski for useful remarks and for comments on the manuscript. We would like to thank Felix Kling for providing us with the high-energy neutrino spectrum relevant for the FASER experiment. KJ is  supported in part by the National Science Centre, Poland, research grant No. 2015/18/A/ST2/00748. ST is supported by the grant ``AstroCeNT: Particle Astrophysics Science and Technology Centre'' carried out within the International Research Agendas programme of the Foundation for Polish Science financed by the European Union under the European Regional Development Fund. ST is supported in part by the Polish Ministry of Science and Higher Education through its scholarship for young and outstanding scientists (decision no 1190/E-78/STYP/14/2019). At the early stage of the work on this project, ST was also supported by the Lancaster-Manchester-Sheffield  Consortium  for Fundamental Physics under STFC grant ST/P000800/1.

\appendix

\section{HNL production in neutrino interactions\label{app:sigma}}

\subsection{Coherent scatterings off nuclei}

Coherent scatterings of the SM neutrinos off nuclei provide typically the dominant secondary production channel for the HNLs in our analysis. This is both due to the $Z^2$ and $(A-Z)^2$ enhancement factors\cite{Freedman:1973yd}, as well as because of typically low momentum transfer associated with such scatterings. This allows to avoid activating veto layers in the secondary production processes happening in front of the detector. In our analysis, we follow Ref.\cite{Jodlowski:2019ycu} for the cuts used in the analysis. In addition, further suppression of the momentum transfer, $|Q^2|<(100~\mev)^2$, is required in our analysis when the scattering events generating high-energy ($E_\gamma>1~\tev$ or $3~\tev$) single photons in the ECC detector are considered. This helps with the identification of the single photon vertices with no additional hadronic activity in the emulsion.

The coherent production cross section of our interest are given by:
\begin{itemize}
\item for the neutrino dipole portal model to HNLs, cf. \cref{sec:dipoleportal} for the coupling constants and model parameters, (see also Ref.\cite{Magill:2018jla,Brdar:2020quo} for the recent discussion)
\begin{align}
\frac{d\sigma(\nu+X^A_Z \to N+X^A_Z)}{dt} &=\frac{-\alpha \mu_N^2 F_1^2(\sqrt{-t})}{t^2(m_T^2-m_N(2E_\nu+m_N))^2} \times\nonumber\\
&\hspace{0.5cm}\times \bigg(m_N t (8 E_\nu^2 m_N + 4 E_\nu (m_N^2 + t) + m_N (m_N^2+t))\nonumber\\
&\hspace{1.2cm} + 2 m_T^2 (-2 m_N t (2 E_\nu + m_N) + m_N^4 - t^2) + 2 m_T^4 t \bigg) \nonumber\\
&+ \frac{\alpha \mu_N^2 F_2^2(\sqrt{-t})}{2 t m_T^2(m_T^2-m_N(2E_\nu+m_N))^2}\times\nonumber\\
&\hspace{0.5cm}\times \bigg(4 t (m_T^2-2 E_\nu m_N) (m_T^2-m_N (2 E_\nu+m_N)) \nonumber\\
&\hspace{1.2cm}+m_N t^2 (8 E_\nu+3 m_N)-4 m_T^2 m_N^4+t^3\bigg),
\label{eq:coherentdipole}
\end{align}
\item for the model with the dark gauge boson, cf. \cref{sec:darkgaugeboson},
\begin{align}
\label{eq:coherentdarkgauge}
\frac{d \sigma(\nu+X_Z^A\to N+X_Z^A)}{d t}&=\frac{\alpha F_1^2(\sqrt{-
t}) |U_{D4}|^2\left(1-|U_{D4}|^2\right)}{32 s_W^4 c_W^4 E_\nu^2 m_T^2
\left(m_Z^2-t\right)^2 \left(m_{Z_D}^2-t\right)^2} \times \\
&\hspace{0.5cm}\times \bigg(8 E_\nu^2 m_T^2+4 E_\nu m_T \left(t-m_N^2\right)-
\left(2m_T^2+t\right) \left(m_N^2-t\right)\bigg) \times \nonumber\\
& \hspace{0.5cm}\times\bigg(\pi\alpha\left(m_{Z_D}^2-t\right)^2 \frac{\left(A+4s_W^2 Z-2 Z \right)^2}{Z^2}\nonumber\\
&\hspace{1.2cm}+32 c_W^4 \epsilon ^2 g_D^2 s_W^4 \left(m_Z^2-t
\right)^2\bigg),\nonumber
\end{align}
\end{itemize}
where $m_T$ and $m_N$ are the masses of the target nucleus and the HNL, respectively, $A$ ($Z$) refers to the atomic mass (number) of the nucleus, and $s_W$ ($c_W$) is the sine (cosine) of the Weinberg angle. In the case of the massive $Z_D$ boson, we take into account the possible additional impact of the scattering mediated by the $Z$ boson from the SM.

In the expressions above, we take the form factor $F_1$ that we parameterize with the Helm form factor $F$\cite{Helm:1956zz} using
\begin{equation}
F_1^2(\sqrt{-t})=Z^2 F^2(\sqrt{-t}),
\end{equation}
where
\begin{equation}
F\equiv F_{\mathrm{Helm}}(Q)=3 \exp\left(\frac{-Q^{2} s^{2}}{2}\right)\frac{\sin (Q r)-Qr \cos (Q r)}{(Q r)^3},
\end{equation}
and $Q=\sqrt{-q^2}=\sqrt{-t}$ is square root of momentum transfer. We put $s=1~\textrm{fm}$, $r=\sqrt{R^2-5s^2}$, and $R=1.2 A^{1/3}~\textrm{fm}$.

We have also included the effect of the screening of the nucleus by the electrons, which could take place in the coherent regime. This is done by multiplying the Helm form factor by the atomic form factor of the form\cite{Kim:1973he,Tsai:1973py}
\begin{equation}
G_{at}(Q)=\frac{a^2 Q^2}{1+a^2Q^2 },
\end{equation}
where $a=111 Z^{-1/3}/m_e$. We have checked this effect has negligible influence on the results. Hence, we do not include it in the formulas for the cross-sections above.

Instead, the magnetic form factor is not known analytically in the coherent regime. However, in this case, there is no enhancement by factor of $Z^2$. Therefore, the relevant contribution to the cross section is strongly suppressed with respect to the one proportional to $F_1$ and we neglect it in our analysis.

\subsection{Elastic incoherent scatterings off individual nucleons}

The neutrino upscattering to HNLs is also possible to take place incoherently off individual protons or neutrons. In this case, the expression for the scattering cross section remains the same as above, cf. \cref{eq:coherentdipole,eq:coherentdarkgauge}, with the target mass replaced by $m_T=m_{p}$ or $m_{n}$, and with different form factors. We take them to be equal to:
\begin{equation}
\begin{aligned}
F_{1}^{p,n} &=\frac{G_{E}^{p,n}+G_{M}^{p,n}\frac{Q^{2}}{4 m_{p}^{2}}}{1+\frac{Q^{2}}{4 m a_{p}^{2}}}, \\
F_{2}^{p,n} &=\frac{G_{M}^{p,n}-G_{E}^{p,n}}{1+\frac{Q^{2}}{4 m_{p}^{2}}}.
\end{aligned}
\end{equation}
In this expression, the electric ($F_{1}^{p,n}$) and magnetic ($F_{2}^{p,n}$) form factors are expressed through the Sachs electric and magnetic form factors $G_{E}^{p,n}$ and $G_{M}^{p,n}$, which in turn are determined experimentally. It was shown that they can be written following the dipole approximation, which remains valid up to $Q^2\sim 10\gev^2$\cite{Qattan:2004ht}. This leads to the following expressions used in our analysis
\begin{equation}
G_{D}=\left(1+\frac{Q^{2}}{0.71 \mathrm{GeV}^{2}}\right)^{-2},
\end{equation}
where
\begin{equation}
\begin{aligned}
G_{E}^{\{p, n\}} &=\left\{G_{D}, 0\right\}, \\
G_{M}^{\{p, n\}} &=\mu_{\{p, n\}} G_{D}, \\
\mu_{p, n} &=\{2.793,-1.913\}.
\end{aligned}
\end{equation}

In the incoherent regime, the total cross-section scales only linearly with the number of nucleons:
\begin{equation}
\sigma_{\text {total,incoh.}}=Z \times \sigma_{p}+(A-Z) \times \sigma_{n}.
\end{equation}

\subsection{Scattering off electrons}

The neutrino upscattering to HNLs is also produced in the interactions with electrons. The relevant cross section formulae read
\begin{itemize}
\item for the neutrino dipole portal model to HNLs,
\begin{equation}
\frac{d\sigma(\nu+e^- \to N+e^-)}{dt} =-\alpha\mu_N^2\frac{2m_e^2(4E_\nu^2 t+m_N^4-m_N^2t)+4E_\nu m_e t(t-m_N^2)+m_N^2t(m_N^2-t)}{2m_e^2E_\nu^2t^2},
\end{equation}
\item for the model with the dark gauge boson,
\begin{align}
\frac{d\sigma(\nu+e^- \to N+e^-)}{dt} &=\alpha \epsilon ^2 g_D^2 U_{4\tau}^2(1-U_{4\tau}^2)\times \nonumber\\
&\frac{\left(8 E_1^2 m_e^2-2 m_N^2 \left(4E_1 m_e+m_e^2+t\right)+2 m_e t (2 E_1+m_e)+2m_N^4+t^2\right)}{2 E_1^2 m_e^2 \left(m_{Z_D}^2-t\right)^2}.
\end{align}
\end{itemize}

In the latter scenario, we also consider elastic scattering off the HNLs off electrons with the relevant cross section given by
\begin{equation}
\frac{d\sigma(N+e^- \to N+e^-)}{dt} =\alpha \epsilon ^2 g_D^2\frac{\left(8 E_1^2 m_e^2+2 t \left(m_e (2E_1+m_e)+m_N^2\right)+t^2\right)}{2 m_e^2 (E_1^2-m_N^2)\left(m_{Z_D}^2-t\right)^2}.
\end{equation}

\section{Decays widths\label{app:decay}}

For completeness, we also provide the relevant decay widths used in our study.

\paragraph{Neutrino dipole portal to HNLs} In the dipole portal scenario, the dominant two-body decay width of the HNL into the SM neutrino and photon is given by\cite{Magill:2018jla}
\begin{equation}
\Gamma=\frac{\mu_N^2 m_N^3}{4\pi}.
\end{equation}

The HNL can also decay into the three-body final state. The corresponding decay width into the pair of leptons and photon, $N\to \gamma\ell\ell$, reads
\begin{equation}
\Gamma_{N\to \gamma\ell\ell}=\frac{1}{512\pi^3 m_N^3} \int_{4m_\ell^2}^{m_N^2}ds_{2}\int_{s_{1}^{min}}^{s_{1}^{max}}ds_{1} |M|^2,
\end{equation}
where
\begin{equation}
|M|^2=\frac{8 \mu_N^2 e^2 \left(2 m_\ell^4 s_2+2 m_\ell^2 \left(m_N^4-s_2 (2 s_1+s_2)\right)+s_2 \left(m_N^4-m_N^2 (2s_1+s_2)+2 s_1 (s_1+s_2)\right)\right)}{s_2^2}.
\end{equation}
The differential cross section can then be written as
\begin{equation}
\frac{d\Gamma_{N\to \gamma\ell\ell}}{ds_2}=\frac{\mu_N^2 e^2 \sqrt{1-\frac{4 m_{\ell}^2}{s_2}} \left(2m_{\ell}^2+s_2\right) \left(m_N^2-s_2\right)^2\left(2 m_N^2+s_2\right)}{192\pi^3 s_2^2 m_N^3}.
\label{eq:diffgamma3body}
\end{equation}
The total cross section reads
\begin{equation}
\Gamma_{N\to \gamma\ell\ell}=\frac{e^2 \mu_N^2\left(\left(8 m_{\ell}^6-2 m_N^6\right) \log \left(\frac{2 m_{\ell}}{\sqrt{m_N^2-4 m_{\ell}^2}+m_N}\right)+m_N \sqrt{m_N^2-4 m_{\ell}^2} \left(-2 m_{\ell}^4+5 m_{\ell}^2 m_N^2-3 m_N^4\right)\right)}{96 \pi ^3 m_N^3}
\end{equation}

When applying the threshold of $100\gev$ for the visible energy in the detector, the integration in \cref{eq:diffgamma3body} needs to be performed with the relevant condition taken into account. We refer to such result as the ``effective'' branching fraction, when discussing the right panel of \cref{fig:gammaspectrum} in \cref{sec:results}.

\paragraph{Model with the dark gauge boson} In the case when $m_N>m_{Z_D}$, in which the decay $N\to Z_D\nu$ is possible, the HNL decay width reads\cite{Bertuzzo:2018itn}
\begin{equation}
\Gamma_{N_{D} \rightarrow Z_{D}+\nu}=\frac{\alpha_{\mathcal{D}}}{2}\left|U_{D 4}\right|^{2}\left(1-\left|U_{D 4}\right|^{2}\right) \frac{m_{N_{\mathcal{D}}}^{3}}{m_{Z_D}^{2}}\left(1-\frac{m_{Z_D}^{2}}{m_{N_{\mathcal{D}}}^{2}}\right)\left(1+\frac{m_{Z_D}^{2}}{m_{N_{\mathcal{D}}}^{2}}-2 \frac{m_{Z_D}^{4}}{m_{N_{\mathcal{D}}}^{4}}\right).
\end{equation}
Here, the dark gauge boson can subsequently decay into the $e^+e^-$ or $\nu\bar{\nu}$ pair with the corresponding decay widths given, respectively, by
\begin{equation}
\Gamma_{Z_{D} \rightarrow e^+e^-} \approx \frac{\alpha \epsilon^2}{3}m_{Z_{D}},
\end{equation}
and
\begin{equation}
\Gamma_{Z_{D} \rightarrow \nu\nu} = \frac{\alpha_D}{3} (1-\left|U_{D 4}\right|^{2})^2 m_{Z_{D}},
\end{equation}
where $\alpha_D=g_D^2/(4\pi)$. In our case, typically, $\alpha \epsilon^{2} \gg \alpha_{\mathcal{D}}\left(1-\left|U_{D 4}\right|^{2}\right)^{2}$ so that $Z_D$ decays mainly into the $e^+e^-$ final state.

In the case when $m_N<m_{Z_D}$, we use result from Ref.\cite{Jho:2020jfz}
\begin{equation}
\Gamma_{N \rightarrow \nu_{\tau} e^{-} e^{+}}=\frac{G_{D}^{2} \epsilon^2}{48 \pi^{3}}\left|U_{\tau 4}\right|^{2} m_N^{5}\left[I_{2}\left(0,\frac{m_e}{m_N},\frac{m_e}{m_N}\right)+2 I_{1}\left(0,\frac{m_e}{m_N},\frac{m_e}{m_N}\right)\right],
\end{equation}
where $G_{D}=g_{D}^{2} /\left(4 \sqrt{2} m_{Z_D}^{2}\right)$ and the functions $I_1$, $I_2$ are defined as follows\cite{Helo:2010cw}:
\begin{equation}
I_{1}(x, y, z)=12 \int_{(x+y)^{2}}^{(1-z)^{2}} \frac{d s}{s}\left(s-x^{2}-y^{2}\right)\left(1+z^{2}-s\right) \lambda^{1 / 2}\left(s, x^{2}, y^{2}\right) \lambda^{1 / 2}\left(1, s, z^{2}\right),
\end{equation}
\begin{equation}
I_{2}(x, y, z)=24 y z \int_{(y+z)^{2}}^{(1-x)^{2}} \frac{d s}{s}\left(1+x^{2}-s\right) \lambda^{1 / 2}\left(s, y^{2}, z^{2}\right) \lambda^{1 / 2}\left(1, s, x^{2}\right),
\end{equation}
where
\begin{equation}
\lambda(x,y,z)=(x-y-z)^2-4yz.
\end{equation}

\bibliographystyle{JHEP}
\bibliography{biblio}

\providecommand{\href}[2]{#2}\begingroup\raggedright\begin{thebibliography}{100}

\bibitem{Fukuda:1998mi}
{\bf Super-Kamiokande} Collaboration, Y.~Fukuda et~al., {\it {Evidence for
  oscillation of atmospheric neutrinos}},  {\em Phys. Rev. Lett.} {\bf 81}
  (1998) 1562--1567, [\href{http://arxiv.org/abs/hep-ex/9807003}{{\tt
  hep-ex/9807003}}].

\bibitem{Ahmad:2001an}
{\bf SNO} Collaboration, Q.~Ahmad et~al., {\it {Measurement of the rate of
  $\nu_e+d \to p+p+e^-$ interactions produced by $^8B$ solar neutrinos at the
  Sudbury Neutrino Observatory}},  {\em Phys. Rev. Lett.} {\bf 87} (2001)
  071301, [\href{http://arxiv.org/abs/nucl-ex/0106015}{{\tt nucl-ex/0106015}}].

\bibitem{Ahmad:2002jz}
{\bf SNO} Collaboration, Q.~Ahmad et~al., {\it {Direct evidence for neutrino
  flavor transformation from neutral current interactions in the Sudbury
  Neutrino Observatory}},  {\em Phys. Rev. Lett.} {\bf 89} (2002) 011301,
  [\href{http://arxiv.org/abs/nucl-ex/0204008}{{\tt nucl-ex/0204008}}].

\bibitem{Aartsen:2013jza}
{\bf IceCube} Collaboration, M.~Aartsen et~al., {\it {Measurement of
  Atmospheric Neutrino Oscillations with IceCube}},  {\em Phys. Rev. Lett.}
  {\bf 111} (2013), no.~8 081801, [\href{http://arxiv.org/abs/1305.3909}{{\tt
  arXiv:1305.3909}}].

\bibitem{Aartsen:2013jdh}
{\bf IceCube} Collaboration, M.~Aartsen et~al., {\it {Evidence for High-Energy
  Extraterrestrial Neutrinos at the IceCube Detector}},  {\em Science} {\bf
  342} (2013) 1242856, [\href{http://arxiv.org/abs/1311.5238}{{\tt
  arXiv:1311.5238}}].

\bibitem{IceCube:2018dnn}
{\bf IceCube, Fermi-LAT, MAGIC, AGILE, ASAS-SN, HAWC, H.E.S.S., INTEGRAL,
  Kanata, Kiso, Kapteyn, Liverpool Telescope, Subaru, Swift NuSTAR, VERITAS,
  VLA/17B-403} Collaboration, M.~Aartsen et~al., {\it {Multimessenger
  observations of a flaring blazar coincident with high-energy neutrino
  IceCube-170922A}},  {\em Science} {\bf 361} (2018), no.~6398 eaat1378,
  [\href{http://arxiv.org/abs/1807.08816}{{\tt arXiv:1807.08816}}].

\bibitem{Zyla:2020zbs}
{\bf Particle Data Group} Collaboration, P.~Zyla et~al., {\it {Review of
  Particle Physics}},  {\em PTEP} {\bf 2020} (2020), no.~8 083C01.

\bibitem{Aartsen:2017kpd}
{\bf IceCube} Collaboration, M.~Aartsen et~al., {\it {Measurement of the
  multi-TeV neutrino cross section with IceCube using Earth absorption}},  {\em
  Nature} {\bf 551} (2017) 596--600,
  [\href{http://arxiv.org/abs/1711.08119}{{\tt arXiv:1711.08119}}].

\bibitem{Bustamante:2017xuy}
M.~Bustamante and A.~Connolly, {\it {Extracting the Energy-Dependent
  Neutrino-Nucleon Cross Section above 10 TeV Using IceCube Showers}},  {\em
  Phys. Rev. Lett.} {\bf 122} (2019), no.~4 041101,
  [\href{http://arxiv.org/abs/1711.11043}{{\tt arXiv:1711.11043}}].

\bibitem{Abreu:2019yak}
{\bf FASER} Collaboration, H.~Abreu et~al., {\it {Detecting and Studying
  High-Energy Collider Neutrinos with FASER at the LHC}},  {\em Eur. Phys. J.
  C} {\bf 80} (2020), no.~1 61, [\href{http://arxiv.org/abs/1908.02310}{{\tt
  arXiv:1908.02310}}].

\bibitem{Abreu:2020ddv}
{\bf FASER} Collaboration, H.~Abreu et~al., {\it {Technical Proposal:
  FASERnu}},  \href{http://arxiv.org/abs/2001.03073}{{\tt arXiv:2001.03073}}.

\bibitem{Jodlowski:2019ycu}
K.~Jod\l{}owski, F.~Kling, L.~Roszkowski, and S.~Trojanowski, {\it {Extending
  the reach of FASER, MATHUSLA, and SHiP towards smaller lifetimes using
  secondary particle production}},  {\em Phys. Rev. D} {\bf 101} (2020), no.~9
  095020, [\href{http://arxiv.org/abs/1911.11346}{{\tt arXiv:1911.11346}}].

\bibitem{Ariga:2018zuc}
{\bf FASER} Collaboration, A.~Ariga et~al., {\it {Letter of Intent for FASER:
  ForwArd Search ExpeRiment at the LHC}},
  \href{http://arxiv.org/abs/1811.10243}{{\tt arXiv:1811.10243}}.

\bibitem{Ariga:2018pin}
{\bf FASER} Collaboration, A.~Ariga et~al., {\it {Technical Proposal for FASER:
  ForwArd Search ExpeRiment at the LHC}},
  \href{http://arxiv.org/abs/1812.09139}{{\tt arXiv:1812.09139}}.

\bibitem{Feng:2017uoz}
J.~L. Feng, I.~Galon, F.~Kling, and S.~Trojanowski, {\it {ForwArd Search
  ExpeRiment at the LHC}},  {\em Phys. Rev. D} {\bf 97} (2018), no.~3 035001,
  [\href{http://arxiv.org/abs/1708.09389}{{\tt arXiv:1708.09389}}].

\bibitem{Ariga:2018uku}
{\bf FASER} Collaboration, A.~Ariga et~al., {\it {FASER\textquoteright{}s
  physics reach for long-lived particles}},  {\em Phys. Rev. D} {\bf 99}
  (2019), no.~9 095011, [\href{http://arxiv.org/abs/1811.12522}{{\tt
  arXiv:1811.12522}}].

\bibitem{Bakhti:2020szu}
P.~Bakhti, Y.~Farzan, and S.~Pascoli, {\it {Discovery potential of FASER$\nu$
  with contained vertex and through-going events}},
  \href{http://arxiv.org/abs/2010.16312}{{\tt arXiv:2010.16312}}.

\bibitem{Minkowski:1977sc}
P.~Minkowski, {\it {$\mu \to e\gamma$ at a Rate of One Out of $10^{9}$ Muon
  Decays?}},  {\em Phys. Lett.} {\bf 67B} (1977) 421--428.

\bibitem{GellMann:1980vs}
M.~Gell-Mann, P.~Ramond, and R.~Slansky, {\it {Complex Spinors and Unified
  Theories}},  {\em Conf. Proc.} {\bf C790927} (1979) 315--321,
  [\href{http://arxiv.org/abs/1306.4669}{{\tt arXiv:1306.4669}}].

\bibitem{Mohapatra:1979ia}
R.~N. Mohapatra and G.~Senjanovic, {\it {Neutrino Mass and Spontaneous Parity
  Violation}},  {\em Phys. Rev. Lett.} {\bf 44} (1980) 912.

\bibitem{Yanagida:1980xy}
T.~Yanagida, {\it {Horizontal Symmetry and Masses of Neutrinos}},  {\em Prog.
  Theor. Phys.} {\bf 64} (1980) 1103.

\bibitem{Schechter:1980gr}
J.~Schechter and J.~W.~F. Valle, {\it {Neutrino Masses in SU(2) x U(1)
  Theories}},  {\em Phys. Rev.} {\bf D22} (1980) 2227.

\bibitem{Drewes:2013gca}
M.~Drewes, {\it {The Phenomenology of Right Handed Neutrinos}},  {\em Int. J.
  Mod. Phys. E} {\bf 22} (2013) 1330019,
  [\href{http://arxiv.org/abs/1303.6912}{{\tt arXiv:1303.6912}}].

\bibitem{Gninenko:2009ks}
S.~Gninenko, {\it {The MiniBooNE anomaly and heavy neutrino decay}},  {\em
  Phys. Rev. Lett.} {\bf 103} (2009) 241802,
  [\href{http://arxiv.org/abs/0902.3802}{{\tt arXiv:0902.3802}}].

\bibitem{Aparici:2009fh}
A.~Aparici, K.~Kim, A.~Santamaria, and J.~Wudka, {\it {Right-handed neutrino
  magnetic moments}},  {\em Phys. Rev. D} {\bf 80} (2009) 013010,
  [\href{http://arxiv.org/abs/0904.3244}{{\tt arXiv:0904.3244}}].

\bibitem{Gninenko:2010pr}
S.~N. Gninenko, {\it {A resolution of puzzles from the LSND, KARMEN, and
  MiniBooNE experiments}},  {\em Phys. Rev. D} {\bf 83} (2011) 015015,
  [\href{http://arxiv.org/abs/1009.5536}{{\tt arXiv:1009.5536}}].

\bibitem{Coloma:2017ppo}
P.~Coloma, P.~A. Machado, I.~Martinez-Soler, and I.~M. Shoemaker, {\it
  {Double-Cascade Events from New Physics in Icecube}},  {\em Phys. Rev. Lett.}
  {\bf 119} (2017), no.~20 201804, [\href{http://arxiv.org/abs/1707.08573}{{\tt
  arXiv:1707.08573}}].

\bibitem{Magill:2018jla}
G.~Magill, R.~Plestid, M.~Pospelov, and Y.-D. Tsai, {\it {Dipole Portal to
  Heavy Neutral Leptons}},  {\em Phys. Rev. D} {\bf 98} (2018), no.~11 115015,
  [\href{http://arxiv.org/abs/1803.03262}{{\tt arXiv:1803.03262}}].

\bibitem{Ballett:2018ynz}
P.~Ballett, S.~Pascoli, and M.~Ross-Lonergan, {\it {U(1)' mediated decays of
  heavy sterile neutrinos in MiniBooNE}},  {\em Phys. Rev. D} {\bf 99} (2019)
  071701, [\href{http://arxiv.org/abs/1808.02915}{{\tt arXiv:1808.02915}}].

\bibitem{Bertuzzo:2018itn}
E.~Bertuzzo, S.~Jana, P.~A. Machado, and R.~Zukanovich~Funchal, {\it {Dark
  Neutrino Portal to Explain MiniBooNE excess}},  {\em Phys. Rev. Lett.} {\bf
  121} (2018), no.~24 241801, [\href{http://arxiv.org/abs/1807.09877}{{\tt
  arXiv:1807.09877}}].

\bibitem{Shoemaker:2018vii}
I.~M. Shoemaker and J.~Wyenberg, {\it {Direct Detection Experiments at the
  Neutrino Dipole Portal Frontier}},  {\em Phys. Rev. D} {\bf 99} (2019), no.~7
  075010, [\href{http://arxiv.org/abs/1811.12435}{{\tt arXiv:1811.12435}}].

\bibitem{Fischer:2019fbw}
O.~Fischer, A.~Hern\'andez-Cabezudo, and T.~Schwetz, {\it {Explaining the
  MiniBooNE excess by a decaying sterile neutrino with mass in the 250 MeV
  range}},  {\em Phys. Rev. D} {\bf 101} (2020), no.~7 075045,
  [\href{http://arxiv.org/abs/1909.09561}{{\tt arXiv:1909.09561}}].

\bibitem{Datta:2020auq}
A.~Datta, S.~Kamali, and D.~Marfatia, {\it {Dark sector origin of the KOTO and
  MiniBooNE anomalies}},  {\em Phys. Lett. B} {\bf 807} (2020) 135579,
  [\href{http://arxiv.org/abs/2005.08920}{{\tt arXiv:2005.08920}}].

\bibitem{Dutta:2020scq}
B.~Dutta, S.~Ghosh, and T.~Li, {\it {Explaining $(g-2)_{\mu,e}$, the KOTO
  anomaly and the MiniBooNE excess in an extended Higgs model with sterile
  neutrinos}},  {\em Phys. Rev. D} {\bf 102} (2020), no.~5 055017,
  [\href{http://arxiv.org/abs/2006.01319}{{\tt arXiv:2006.01319}}].

\bibitem{Abdallah:2020biq}
W.~Abdallah, R.~Gandhi, and S.~Roy, {\it {Understanding the MiniBooNE and the
  muon $g-2$ anomalies with a light $Z'$ and a second Higgs doublet}},
  \href{http://arxiv.org/abs/2006.01948}{{\tt arXiv:2006.01948}}.

\bibitem{Abdullahi:2020nyr}
A.~Abdullahi, M.~Hostert, and S.~Pascoli, {\it {A Dark Seesaw Solution to Low
  Energy Anomalies: MiniBooNE, the muon $(g-2)$, and BaBar}},
  \href{http://arxiv.org/abs/2007.11813}{{\tt arXiv:2007.11813}}.

\bibitem{Shoemaker:2020kji}
I.~M. Shoemaker, Y.-D. Tsai, and J.~Wyenberg, {\it {An Active-to-Sterile
  Neutrino Transition Dipole Moment and the XENON1T Excess}},
  \href{http://arxiv.org/abs/2007.05513}{{\tt arXiv:2007.05513}}.

\bibitem{Brdar:2020quo}
V.~Brdar, A.~Greljo, J.~Kopp, and T.~Opferkuch, {\it {The Neutrino Magnetic
  Moment Portal: Cosmology, Astrophysics, and Direct Detection}},
  \href{http://arxiv.org/abs/2007.15563}{{\tt arXiv:2007.15563}}.

\bibitem{Jho:2020jfz}
Y.~Jho, J.~Kim, P.~Ko, and S.~C. Park, {\it {Search for sterile neutrino with
  light gauge interactions: recasting collider, beam-dump, and neutrino
  telescope searches}},  \href{http://arxiv.org/abs/2008.12598}{{\tt
  arXiv:2008.12598}}.

\bibitem{Plestid:2020vqf}
R.~Plestid, {\it {Luminous solar neutrinos I: Dipole portals}},
  \href{http://arxiv.org/abs/2010.04193}{{\tt arXiv:2010.04193}}.

\bibitem{Lindner:2018kjo}
M.~Lindner, F.~S. Queiroz, W.~Rodejohann, and X.-J. Xu, {\it {Neutrino-electron
  scattering: general constraints on Z$^{\prime}$ and dark photon models}},
  {\em JHEP} {\bf 05} (2018) 098, [\href{http://arxiv.org/abs/1803.00060}{{\tt
  arXiv:1803.00060}}].

\bibitem{Lindner:2016wff}
M.~Lindner, W.~Rodejohann, and X.-J. Xu, {\it {Coherent Neutrino-Nucleus
  Scattering and new Neutrino Interactions}},  {\em JHEP} {\bf 03} (2017) 097,
  [\href{http://arxiv.org/abs/1612.04150}{{\tt arXiv:1612.04150}}].

\bibitem{Beacham:2019nyx}
J.~Beacham et~al., {\it {Physics Beyond Colliders at CERN: Beyond the Standard
  Model Working Group Report}},  {\em J. Phys. G} {\bf 47} (2020), no.~1
  010501, [\href{http://arxiv.org/abs/1901.09966}{{\tt arXiv:1901.09966}}].

\bibitem{Alimena:2019zri}
J.~Alimena et~al., {\it {Searching for long-lived particles beyond the Standard
  Model at the Large Hadron Collider}},  {\em J. Phys. G} {\bf 47} (2020),
  no.~9 090501, [\href{http://arxiv.org/abs/1903.04497}{{\tt
  arXiv:1903.04497}}].

\bibitem{Agrawal:2021dbo}
P.~Agrawal et~al., {\it {Feebly-Interacting Particles:FIPs 2020 Workshop
  Report}},  \href{http://arxiv.org/abs/2102.12143}{{\tt arXiv:2102.12143}}.

\bibitem{Ariga:2020lbq}
A.~Ariga, T.~Ariga, G.~De~Lellis, A.~Ereditato, and K.~Niwa, {\em {Nuclear
  Emulsions}}.
\newblock 2020.
\newblock {Particle Physics Reference Library}: {Volume 2: Detectors for
  Particles and Radiation}.

\bibitem{Batell:2021blf}
B.~Batell, J.~L. Feng, and S.~Trojanowski, {\it {Detecting Dark Matter with
  Far-Forward Emulsion and Liquid Argon Detectors at the LHC}},
  \href{http://arxiv.org/abs/2101.10338}{{\tt arXiv:2101.10338}}.

\bibitem{Beni:2020yfy}
N.~Beni et~al., {\it {Further studies on the physics potential of an experiment
  using LHC neutrinos}},  \href{http://arxiv.org/abs/2004.07828}{{\tt
  arXiv:2004.07828}}.

\bibitem{CRMC}
C.~Baus, T.~Pierog, and R.~Ulrich, {\it {Cosmic Ray Monte Carlo (CRMC)}}, .
  \url{https://web.ikp.kit.edu/rulrich/crmc.html}.

\bibitem{Pierog:2013ria}
T.~Pierog, I.~Karpenko, J.~M. Katzy, E.~Yatsenko, and K.~Werner, {\it {EPOS
  LHC: Test of collective hadronization with data measured at the CERN Large
  Hadron Collider}},  {\em Phys. Rev.} {\bf C92} (2015), no.~3 034906,
  [\href{http://arxiv.org/abs/1306.0121}{{\tt arXiv:1306.0121}}].

\bibitem{Blumlein:2013cua}
J.~Bl\"umlein and J.~Brunner, {\it {New Exclusion Limits on Dark Gauge Forces
  from Proton Bremsstrahlung in Beam-Dump Data}},  {\em Phys. Lett. B} {\bf
  731} (2014) 320--326, [\href{http://arxiv.org/abs/1311.3870}{{\tt
  arXiv:1311.3870}}].

\bibitem{deNiverville:2016rqh}
P.~deNiverville, C.-Y. Chen, M.~Pospelov, and A.~Ritz, {\it {Light dark matter
  in neutrino beams: production modelling and scattering signatures at
  MiniBooNE, T2K and SHiP}},  {\em Phys. Rev. D} {\bf 95} (2017), no.~3 035006,
  [\href{http://arxiv.org/abs/1609.01770}{{\tt arXiv:1609.01770}}].

\bibitem{Kling:2020mch}
F.~Kling and S.~Trojanowski, {\it {Looking forward to test the KOTO anomaly
  with FASER}},  {\em Phys. Rev. D} {\bf 102} (2020), no.~1 015032,
  [\href{http://arxiv.org/abs/2006.10630}{{\tt arXiv:2006.10630}}].

\bibitem{Ahdida:2020evc}
{\bf SHiP} Collaboration, C.~Ahdida et~al., {\it {SND@LHC}},
  \href{http://arxiv.org/abs/2002.08722}{{\tt arXiv:2002.08722}}.

\bibitem{Ferrari:2005zk}
A.~Ferrari, P.~R. Sala, A.~Fasso, and J.~Ranft, {\it {FLUKA: A multi-particle
  transport code (Program version 2005)}}, .

\bibitem{Battistoni:2015epi}
G.~Battistoni et~al., {\it {Overview of the FLUKA code}},  {\em Annals Nucl.
  Energy} {\bf 82} (2015) 10--18.

\bibitem{FLUKAstudy}
{CERN Sources, Targets, and Interactions Group}, M.~Sabate-Gilarte, F.~Cerutti,
  and A.~Tsinganis, {\it {Characterization of the radiation field for the FASER
  experiment}}, .

\bibitem{Andreopoulos:2009rq}
C.~Andreopoulos et~al., {\it {The GENIE Neutrino Monte Carlo Generator}},  {\em
  Nucl. Instrum. Meth. A} {\bf 614} (2010) 87--104,
  [\href{http://arxiv.org/abs/0905.2517}{{\tt arXiv:0905.2517}}].

\bibitem{Andreopoulos:2015wxa}
C.~Andreopoulos, C.~Barry, S.~Dytman, H.~Gallagher, T.~Golan, R.~Hatcher,
  G.~Perdue, and J.~Yarba, {\it {The GENIE Neutrino Monte Carlo Generator:
  Physics and User Manual}},  \href{http://arxiv.org/abs/1510.05494}{{\tt
  arXiv:1510.05494}}.

\bibitem{Kling:2018wct}
F.~Kling and S.~Trojanowski, {\it {Heavy Neutral Leptons at FASER}},  {\em
  Phys. Rev. D} {\bf 97} (2018), no.~9 095016,
  [\href{http://arxiv.org/abs/1801.08947}{{\tt arXiv:1801.08947}}].

\bibitem{Helo:2018qej}
J.~C. Helo, M.~Hirsch, and Z.~S. Wang, {\it {Heavy neutral fermions at the
  high-luminosity LHC}},  {\em JHEP} {\bf 07} (2018) 056,
  [\href{http://arxiv.org/abs/1803.02212}{{\tt arXiv:1803.02212}}].

\bibitem{Farzan:2018gtr}
Y.~Farzan, M.~Lindner, W.~Rodejohann, and X.-J. Xu, {\it {Probing neutrino
  coupling to a light scalar with coherent neutrino scattering}},  {\em JHEP}
  {\bf 05} (2018) 066, [\href{http://arxiv.org/abs/1802.05171}{{\tt
  arXiv:1802.05171}}].

\bibitem{Petcov:1976ff}
S.~Petcov, {\it {The Processes mu --\ensuremath{>} e Gamma, mu --\ensuremath{>}
  e e anti-e, Neutrino' --\ensuremath{>} Neutrino gamma in the Weinberg-Salam
  Model with Neutrino Mixing}},  {\em Sov. J. Nucl. Phys.} {\bf 25} (1977) 340.
  [Erratum: Sov.J.Nucl.Phys. 25, 698 (1977), Erratum: Yad.Fiz. 25, 1336
  (1977)].

\bibitem{Fujikawa:1980yx}
K.~Fujikawa and R.~Shrock, {\it {The Magnetic Moment of a Massive Neutrino and
  Neutrino Spin Rotation}},  {\em Phys. Rev. Lett.} {\bf 45} (1980) 963.

\bibitem{Pal:1981rm}
P.~B. Pal and L.~Wolfenstein, {\it {Radiative Decays of Massive Neutrinos}},
  {\em Phys. Rev. D} {\bf 25} (1982) 766.

\bibitem{Shrock:1982sc}
R.~E. Shrock, {\it {Electromagnetic Properties and Decays of Dirac and Majorana
  Neutrinos in a General Class of Gauge Theories}},  {\em Nucl. Phys. B} {\bf
  206} (1982) 359--379.

\bibitem{Dvornikov:2003js}
M.~Dvornikov and A.~Studenikin, {\it {Electric charge and magnetic moment of
  massive neutrino}},  {\em Phys. Rev. D} {\bf 69} (2004) 073001,
  [\href{http://arxiv.org/abs/hep-ph/0305206}{{\tt hep-ph/0305206}}].

\bibitem{Giunti:2014ixa}
C.~Giunti and A.~Studenikin, {\it {Neutrino electromagnetic interactions: a
  window to new physics}},  {\em Rev. Mod. Phys.} {\bf 87} (2015) 531,
  [\href{http://arxiv.org/abs/1403.6344}{{\tt arXiv:1403.6344}}].

\bibitem{Lindner:2017uvt}
M.~Lindner, B.~Radov\v{c}i\'c, and J.~Welter, {\it {Revisiting Large Neutrino
  Magnetic Moments}},  {\em JHEP} {\bf 07} (2017) 139,
  [\href{http://arxiv.org/abs/1706.02555}{{\tt arXiv:1706.02555}}].

\bibitem{Babu:2020ivd}
K.~Babu, S.~Jana, and M.~Lindner, {\it {Large Neutrino Magnetic Moments in the
  Light of Recent Experiments}},  \href{http://arxiv.org/abs/2007.04291}{{\tt
  arXiv:2007.04291}}.

\bibitem{AguilarArevalo:2007it}
{\bf MiniBooNE} Collaboration, A.~Aguilar-Arevalo et~al., {\it {A Search for
  Electron Neutrino Appearance at the $\Delta m^2 \sim 1 eV^2$ Scale}},  {\em
  Phys. Rev. Lett.} {\bf 98} (2007) 231801,
  [\href{http://arxiv.org/abs/0704.1500}{{\tt arXiv:0704.1500}}].

\bibitem{Athanassopoulos:1996jb}
{\bf LSND} Collaboration, C.~Athanassopoulos et~al., {\it {Evidence for
  anti-muon-neutrino ---> anti-electron-neutrino oscillations from the LSND
  experiment at LAMPF}},  {\em Phys. Rev. Lett.} {\bf 77} (1996) 3082--3085,
  [\href{http://arxiv.org/abs/nucl-ex/9605003}{{\tt nucl-ex/9605003}}].

\bibitem{Aguilar-Arevalo:2020nvw}
{\bf MiniBooNE} Collaboration, A.~Aguilar-Arevalo et~al., {\it {Updated
  MiniBooNE Neutrino Oscillation Results with Increased Data and New Background
  Studies}},  \href{http://arxiv.org/abs/2006.16883}{{\tt arXiv:2006.16883}}.

\bibitem{Brdar:2020tle}
V.~Brdar, O.~Fischer, and A.~Y. Smirnov, {\it {Model Independent Bounds on the
  Non-Oscillatory Explanations of the MiniBooNE Excess}},
  \href{http://arxiv.org/abs/2007.14411}{{\tt arXiv:2007.14411}}.

\bibitem{Park:2015eqa}
{\bf MINERvA} Collaboration, J.~Park et~al., {\it {Measurement of Neutrino Flux
  from Neutrino-Electron Elastic Scattering}},  {\em Phys. Rev. D} {\bf 93}
  (2016), no.~11 112007, [\href{http://arxiv.org/abs/1512.07699}{{\tt
  arXiv:1512.07699}}].

\bibitem{Jordan:2018qiy}
J.~R. Jordan, Y.~Kahn, G.~Krnjaic, M.~Moschella, and J.~Spitz, {\it {Severe
  Constraints on New Physics Explanations of the MiniBooNE Excess}},  {\em
  Phys. Rev. Lett.} {\bf 122} (2019), no.~8 081801,
  [\href{http://arxiv.org/abs/1810.07185}{{\tt arXiv:1810.07185}}].

\bibitem{Bauer:2018onh}
M.~Bauer, P.~Foldenauer, and J.~Jaeckel, {\it {Hunting All the Hidden
  Photons}},  {\em JHEP} {\bf 18} (2020) 094,
  [\href{http://arxiv.org/abs/1803.05466}{{\tt arXiv:1803.05466}}].

\bibitem{Kling:2020iar}
F.~Kling, {\it {Probing light gauge bosons in tau neutrino experiments}},  {\em
  Phys. Rev. D} {\bf 102} (2020), no.~1 015007,
  [\href{http://arxiv.org/abs/2005.03594}{{\tt arXiv:2005.03594}}].

\bibitem{Orloff:2002de}
J.~Orloff, A.~N. Rozanov, and C.~Santoni, {\it {Limits on the mixing of tau
  neutrino to heavy neutrinos}},  {\em Phys. Lett. B} {\bf 550} (2002) 8--15,
  [\href{http://arxiv.org/abs/hep-ph/0208075}{{\tt hep-ph/0208075}}].

\bibitem{Abe:2019kgx}
{\bf T2K} Collaboration, K.~Abe et~al., {\it {Search for heavy neutrinos with
  the T2K near detector ND280}},  {\em Phys. Rev. D} {\bf 100} (2019), no.~5
  052006, [\href{http://arxiv.org/abs/1902.07598}{{\tt arXiv:1902.07598}}].

\bibitem{Vilain:1994qy}
{\bf CHARM-II} Collaboration, P.~Vilain et~al., {\it {Precision measurement of
  electroweak parameters from the scattering of muon-neutrinos on electrons}},
  {\em Phys. Lett. B} {\bf 335} (1994) 246--252.

\bibitem{Arguelles:2018mtc}
C.~A. Arg\"uelles, M.~Hostert, and Y.-D. Tsai, {\it {Testing New Physics
  Explanations of the MiniBooNE Anomaly at Neutrino Scattering Experiments}},
  {\em Phys. Rev. Lett.} {\bf 123} (2019), no.~26 261801,
  [\href{http://arxiv.org/abs/1812.08768}{{\tt arXiv:1812.08768}}].

\bibitem{Geiregat:1989sz}
{\bf CHARM-II} Collaboration, D.~Geiregat et~al., {\it {A New Determination of
  the Electroweak Mixing Angle From $\nu_\mu$ Electron Scattering}},  {\em
  Phys. Lett. B} {\bf 232} (1989) 539.

\bibitem{Altegoer:1997gv}
{\bf NOMAD} Collaboration, J.~Altegoer et~al., {\it {The NOMAD experiment at
  the CERN SPS}},  {\em Nucl. Instrum. Meth. A} {\bf 404} (1998) 96--128.

\bibitem{Abreu:1996vd}
{\bf DELPHI} Collaboration, P.~Abreu et~al., {\it {Search for new phenomena
  using single photon events in the DELPHI detector at LEP}},  {\em Z. Phys. C}
  {\bf 74} (1997) 577--586.

\bibitem{Groom:2001kq}
D.~E. Groom, N.~V. Mokhov, and S.~I. Striganov, {\it {Muon stopping power and
  range tables 10-MeV to 100-TeV}},  {\em Atom. Data Nucl. Data Tabl.} {\bf 78}
  (2001) 183--356.

\bibitem{Atre:2009rg}
A.~Atre, T.~Han, S.~Pascoli, and B.~Zhang, {\it {The Search for Heavy Majorana
  Neutrinos}},  {\em JHEP} {\bf 05} (2009) 030,
  [\href{http://arxiv.org/abs/0901.3589}{{\tt arXiv:0901.3589}}].

\bibitem{deGouvea:2015euy}
A.~de~Gouv\^ea and A.~Kobach, {\it {Global Constraints on a Heavy Neutrino}},
  {\em Phys. Rev. D} {\bf 93} (2016), no.~3 033005,
  [\href{http://arxiv.org/abs/1511.00683}{{\tt arXiv:1511.00683}}].

\bibitem{Bolton:2019pcu}
P.~D. Bolton, F.~F. Deppisch, and P.~Bhupal~Dev, {\it {Neutrinoless double beta
  decay versus other probes of heavy sterile neutrinos}},  {\em JHEP} {\bf 03}
  (2020) 170, [\href{http://arxiv.org/abs/1912.03058}{{\tt arXiv:1912.03058}}].

\bibitem{Ballett:2019pyw}
P.~Ballett, M.~Hostert, and S.~Pascoli, {\it {Dark Neutrinos and a Three Portal
  Connection to the Standard Model}},  {\em Phys. Rev. D} {\bf 101} (2020),
  no.~11 115025, [\href{http://arxiv.org/abs/1903.07589}{{\tt
  arXiv:1903.07589}}].

\bibitem{Philcox:2020vvt}
O.~H.~E. Philcox, M.~M. Ivanov, M.~Simonovi\'c, and M.~Zaldarriaga, {\it
  {Combining Full-Shape and BAO Analyses of Galaxy Power Spectra: A
  1.6\textbackslash{}\% CMB-independent constraint on H$_0$}},  {\em JCAP} {\bf
  05} (2020) 032, [\href{http://arxiv.org/abs/2002.04035}{{\tt
  arXiv:2002.04035}}].

\bibitem{Bernal:2016gxb}
J.~L. Bernal, L.~Verde, and A.~G. Riess, {\it {The trouble with $H_0$}},  {\em
  JCAP} {\bf 10} (2016) 019, [\href{http://arxiv.org/abs/1607.05617}{{\tt
  arXiv:1607.05617}}].

\bibitem{Knox:2019rjx}
L.~Knox and M.~Millea, {\it {Hubble constant hunter\textquoteright{}s guide}},
  {\em Phys. Rev. D} {\bf 101} (2020), no.~4 043533,
  [\href{http://arxiv.org/abs/1908.03663}{{\tt arXiv:1908.03663}}].

\bibitem{Kobach:2014hea}
A.~Kobach and S.~Dobbs, {\it {Heavy Neutrinos and the Kinematics of Tau
  Decays}},  {\em Phys. Rev. D} {\bf 91} (2015), no.~5 053006,
  [\href{http://arxiv.org/abs/1412.4785}{{\tt arXiv:1412.4785}}].

\bibitem{Abreu:1996pa}
{\bf DELPHI} Collaboration, P.~Abreu et~al., {\it {Search for neutral heavy
  leptons produced in Z decays}},  {\em Z. Phys. C} {\bf 74} (1997) 57--71.
  [Erratum: Z.Phys.C 75, 580 (1997)].

\bibitem{Astier:2001ck}
{\bf NOMAD} Collaboration, P.~Astier et~al., {\it {Search for heavy neutrinos
  mixing with tau neutrinos}},  {\em Phys. Lett. B} {\bf 506} (2001) 27--38,
  [\href{http://arxiv.org/abs/hep-ex/0101041}{{\tt hep-ex/0101041}}].

\bibitem{Geiregat:1991md}
{\bf CHARM-II} Collaboration, D.~Geiregat et~al., {\it {An Improved
  determination of the electroweak mixing angle from muon-neutrino electron
  scattering}},  {\em Phys. Lett. B} {\bf 259} (1991) 499--507.

\bibitem{Vilain:1992wx}
{\bf CHARM-II} Collaboration, P.~Vilain et~al., {\it {Neutral current coupling
  constants from neutrino and anti-neutrino - electron scattering}},  {\em
  Phys. Lett. B} {\bf 281} (1992) 159--166.

\bibitem{Foroughi-Abari:2020qar}
S.~Foroughi-Abari, F.~Kling, and Y.-D. Tsai, {\it {FORMOSA: Looking Forward to
  Millicharged Dark Sectors}},  \href{http://arxiv.org/abs/2010.07941}{{\tt
  arXiv:2010.07941}}.

\bibitem{Altmannshofer:2014pba}
W.~Altmannshofer, S.~Gori, M.~Pospelov, and I.~Yavin, {\it {Neutrino Trident
  Production: A Powerful Probe of New Physics with Neutrino Beams}},  {\em
  Phys. Rev. Lett.} {\bf 113} (2014) 091801,
  [\href{http://arxiv.org/abs/1406.2332}{{\tt arXiv:1406.2332}}].

\bibitem{SnowmassFPF}
R.~M. Abraham et~al., {\em Forward Physics Facility: Snowmass 2021 Letter of
  Interest}, Aug., 2020.

\bibitem{Freedman:1973yd}
D.~Z. Freedman, {\it {Coherent Neutrino Nucleus Scattering as a Probe of the
  Weak Neutral Current}},  {\em Phys. Rev. D} {\bf 9} (1974) 1389--1392.

\bibitem{Helm:1956zz}
R.~H. Helm, {\it {Inelastic and Elastic Scattering of 187-Mev Electrons from
  Selected Even-Even Nuclei}},  {\em Phys. Rev.} {\bf 104} (1956) 1466--1475.

\bibitem{Kim:1973he}
K.~J. Kim and Y.-S. Tsai, {\it {Improved Weizsacker-Williams method and its
  application to lepton and W boson pair production}},  {\em Phys. Rev.} {\bf
  D8} (1973) 3109.

\bibitem{Tsai:1973py}
Y.-S. Tsai, {\it {Pair Production and Bremsstrahlung of Charged Leptons}},
  {\em Rev. Mod. Phys.} {\bf 46} (1974) 815. [Erratum: Rev. Mod.
  Phys.49,521(1977)].

\bibitem{Qattan:2004ht}
I.~Qattan et~al., {\it {Precision Rosenbluth measurement of the proton elastic
  form-factors}},  {\em Phys. Rev. Lett.} {\bf 94} (2005) 142301,
  [\href{http://arxiv.org/abs/nucl-ex/0410010}{{\tt nucl-ex/0410010}}].

\bibitem{Helo:2010cw}
J.~C. Helo, S.~Kovalenko, and I.~Schmidt, {\it {Sterile neutrinos in lepton
  number and lepton flavor violating decays}},  {\em Nucl. Phys. B} {\bf 853}
  (2011) 80--104, [\href{http://arxiv.org/abs/1005.1607}{{\tt
  arXiv:1005.1607}}].

\end{thebibliography}\endgroup

\end{document}